%% file: main.tex
\def\isarxiv{1} 

\ifdefined\isarxiv
\documentclass[11pt]{article}

\usepackage[numbers]{natbib}

\else
\documentclass{article}
\usepackage{neurips_2022}
\fi

\usepackage{amsmath}
\usepackage{amsthm}
\usepackage{amssymb}
\usepackage{algorithm}
\usepackage{subfig}
\usepackage{algpseudocode}
\usepackage{graphicx}
\usepackage{grffile}
\usepackage{wrapfig,epsfig}
\usepackage{url}
\usepackage{xcolor}
\usepackage{epstopdf}

\usepackage{bbm}
\usepackage{dsfont}

\allowdisplaybreaks

\ifdefined\isarxiv

\usepackage{tikz}
\usepackage{hyperref}  
\hypersetup{colorlinks=true,citecolor=blue,linkcolor=blue} 
\usetikzlibrary{arrows}
\usepackage[margin=1in]{geometry}

\else

\usepackage{microtype}
\usepackage{hyperref}
\definecolor{mydarkblue}{rgb}{0,0.08,0.45}
\hypersetup{colorlinks=true, citecolor=mydarkblue,linkcolor=mydarkblue}

\fi

\newtheorem{theorem}{Theorem}[section]
\newtheorem{lemma}[theorem]{Lemma}
\newtheorem{definition}[theorem]{Definition}

\newtheorem{fact}[theorem]{Fact}

\newtheorem{claim}[theorem]{Claim}

\newcommand{\wh}{\widehat}
\newcommand{\wt}{\widetilde}
\newcommand{\ov}{\overline}
\newcommand{\N}{\mathcal{N}}
\newcommand{\R}{\mathbb{R}}

\renewcommand{\hat}{\wh}
\newcommand{\Tmat}{{\cal T}_{\mathrm{mat}}}

\DeclareMathOperator*{\E}{{\mathbb{E}}}

\DeclareMathOperator{\poly}{poly}

\DeclareMathOperator{\rank}{rank}
\DeclareMathOperator{\dist}{dist}

\DeclareMathOperator{\vect}{vec}
\DeclareMathOperator{\tr}{tr}

\DeclareMathOperator{\Var}{Var}

\makeatletter
\newcommand*{\RN}[1]{\expandafter\@slowromancap\romannumeral #1@}
\makeatother

\usepackage{lineno}

\begin{document}

\ifdefined\isarxiv

\date{}

\title{An Improved Sample Complexity for Rank-1 Matrix Sensing}
\author{
Yichuan Deng\thanks{\texttt{ethandeng02@gmail.com}. University of Science and Technology of China.}
\and 
Zhihang Li\thanks{\texttt{lizhihangdll@gmail.com}. Huazhong Agriculture University.} 
\and
Zhao Song\thanks{\texttt{zsong@adobe.com}. Adobe Research}
}

\else

\title{Intern Project} 
\maketitle 
\fi

\ifdefined\isarxiv
\begin{titlepage}
  \maketitle
  \begin{abstract}
\input{abstract}

  \end{abstract}
  \thispagestyle{empty}
\end{titlepage}

\newpage

\else

\begin{abstract}
\input{abstract}
\end{abstract}

\fi

\input{intro} 

\input{related}

\input{tech}

\input{preli}

\input{analysis}

\input{shrink}

\input{reg}

\ifdefined\isarxiv
\bibliographystyle{alpha}
\bibliography{ref}
\else
\bibliography{ref}
\bibliographystyle{alpha}

\fi

\newpage
\onecolumn
\appendix




\end{document}

%% file: abstract.tex
Matrix sensing is a problem in signal processing and machine learning that involves recovering a low-rank matrix from a set of linear measurements. The goal is to reconstruct the original matrix as accurately as possible, given only a set of linear measurements obtained by sensing the matrix \cite{jns13}. In this work, we focus on a particular direction of matrix sensing, which is called rank-$1$ matrix sensing \cite{zjd15}. 
We present an improvement over the original algorithm in \cite{zjd15}. 
It is based on a novel analysis and sketching technique that enables faster convergence rates and better accuracy in recovering low-rank matrices. The algorithm focuses on developing a theoretical understanding of the matrix sensing problem and establishing its advantages over previous methods. The proposed sketching technique allows for efficiently extracting relevant information from the linear measurements, making the algorithm computationally efficient and scalable. 

Our novel matrix sensing algorithm improves former result \cite{zjd15} on in two senses,
\begin{itemize}
    \item We improve the sample complexity from $\wt{O}(\epsilon^{-2} dk^2)$ to $\wt{O}(\epsilon^{-2} (d+k^2))$.
    \item We improve the running time from $\wt{O}(md^2 k^2)$ to $\wt{O}(m d^2 k)$.   
\end{itemize}
The proposed algorithm has theoretical guarantees and is analyzed to provide insights into the underlying structure of low-rank matrices and the nature of the linear measurements used in the recovery process. 
It advances the theoretical understanding of matrix sensing and provides a new approach for solving this important problem. 

%% file: intro.tex
\section{Introduction}

The matrix sensing problem is a fundamental problem in signal processing and machine learning that involves recovering a low-rank matrix from a set of linear measurement. This problem arises in various applications such as image and video processing \cite{fmt+12, bjz+18} and sensor networks \cite{mcn17, wv17}. 
Mathematically, matrix sensing can be formulated as a matrix view of compressive sensing problem \cite{jns13}. The rank-$1$ matrix sensing problem was formally raised in \cite{zjd15}. 

The matrix sensing problem has attracted significant attention in recent years, and several algorithms have been proposed to solve it efficiently. In this paper, we provide a novel improvement over the origin algorithm in \cite{zjd15}, with improvement both on running time and sample complexity.

Matrix sensing is a fundamental problem in signal processing and machine learning that involves recovering a low-rank matrix from a set of linear measurements. Specifically, given a matrix $W_*\in\mathbb{R}^{d \times d}$ of rank $k$ that is not directly accessible, we aim to recover $W_*$ from a set of linear measurements $b \in \R^n$ applied to the ground truth matrix $W^*$ where
\begin{align*}
    b_i=\tr[A_i^\top W_*],~~~\forall i=1, \dots, m, 
\end{align*}
where $A_i$ are known linear operators. The measurements $b_i$ are obtained by sensing the matrix $W_*$ using a set of linear measurements, and the goal is to reconstruct the original matrix $W_*$ as accurately as possible. This problem arises in various applications such as image and video processing, sensor networks, and recommendation systems.

The matrix sensing problem is ill-posed since there may exist multiple low-rank matrices that satisfy the given linear measurements. However, the problem becomes well-posed under some assumptions on the underlying matrix, such as incoherence and restricted isometry property (RIP) \cite{ct05, crt06, gh08}
, which ensure unique and stable recovery of the matrix. A well-used method to solve this problem is to use convex optimization techniques that minimize a certain loss function subject to the linear constraints. Specifically, one can solve the following convex optimization problem:
\begin{align*}
        &~\min_{W_*}~\rank(W_*)\\
    \text{s.t.}&~\tr[A_i^\top W_*] = b_i, \forall i=1,\dots,m.
\end{align*}
However, this problem is NP-hard \cite{tp13} and intractable in general, and hence, various relaxation methods 
have been proposed, such as nuclear norm minimization and its variants, which provide computationally efficient solutions with theoretical guarantees. In this work, we focus on the \emph{rank-one independent} measurements. Under this setting, the linear operators $A_i$ can be decomposed into the form of $A_i = x_iy_i^\top$, where $x_i \in \R^{d}, y_i \in \R^{d}$ are all sampled from zero-mean multivariate Gaussian distribution ${\cal N}(0, I_d)$. 

Our work on improving the matrix sensing algorithm is based on a novel analysis and sketching technique 
that enables faster convergence rates and better accuracy in recovering low-rank matrices. We focus on developing a theoretical understanding of the proposed algorithm and establishing its advantages over previous methods. Our analysis provides insights into the underlying structure of the low-rank matrices and the nature of the linear measurements used in the recovery process. The proposed sketching technique allows us to efficiently extract relevant information from the linear measurements, making our algorithm computationally efficient and scalable. Overall, our contribution advances the theoretical understanding of matrix sensing and provides a new approach for solving this important problem.

\subsection{Our Result}
To summarize, we improve both the running time of original algorithm \cite{zjd15} from $O(md^2k^2)$ to $O(md^2k)$, and the sample complexity from $\wt{O}(\epsilon^{-2}dk^2)$ to $\wt{O}(\epsilon^{-2}(d + k^2))$. Formally, we get the following result,

\begin{theorem}[Informal, combination of Theorem~\ref{thm:main_convergence}, Theorem~\ref{thm:main_measurement} and Theorem~\ref{thm:main_cost}]\label{thm:main_informal}
Let $\epsilon_0 \in (0,0.1)$ denote the final accuracy of the algorithm. Let $\delta \in (0,0.1)$ denote the failure probability of the algorithm. Let $\sigma_1^*$ denote the largest singular value of ground-truth matrix $W_* \in \R^{d \times d}$. Let $\kappa$ denote the condition number of ground-truth matrix $W_* \in \R^{d \times d}$. 
Let $\epsilon \in (0, 0.001/(k^{1.5}\kappa))$ denote the RIP parameter. Let $m = \Theta (\epsilon^{-2} (d+k^2)\log(d/\delta))$. Let $T = \Theta(\log(k \kappa \sigma_1^* /\epsilon_0))$ . There is a matrix sensing algorithm (Algorithm~\ref{alg:main}) that takes $O(m T)$ samples, runs in $T$ iterations, and each iteration takes $\wt{O}(md^2 k)$ time, finally outputs a matrix $W \in \R^{d \times d}$ such that
\begin{align*}
 (1-\epsilon_0) W_* \preceq W \preceq (1+\epsilon_0) W_*
\end{align*}
holds with probability at least $1-\delta$.
\end{theorem}

%% file: related.tex
\subsection{Related Work}

\paragraph{Matrix Sensing}

The matrix sensing problem has attracted significant attention in recent years, and several algorithms have been proposed to solve it efficiently. One of the earliest approaches is the convex optimization-based algorithm proposed by Candès and Recht in 2009 \cite{cr12}, which minimizes the nuclear norm of the matrix subject to the linear constraints. This approach has been shown to achieve optimal recovery guarantees under certain conditions on the linear operators, such as incoherence and RIP.
Since then, various algorithms have been proposed that improve upon the original approach in terms of computational efficiency and theoretical guarantees. For instance, the iterative hard
thresholding algorithm (IHT) proposed by Blumensath and Davies in 2009 \cite{bd08}, and its variants, such
as the iterative soft thresholding algorithm (IST), provide computationally efficient solutions with
improved recovery guarantees. 
In the work by Recht, Fazel, and Parrilo 
\cite{rfp10}, they gave some measurement operators satisfying the RIP and proved that, with $O(k d \log d)$ measurements, a rank-$k$ matrix $W_* \in \R^{d \times d}$ can be recovered. 
Moreover, later works have proposed new approaches that exploit additional structure in the low-rank matrix, such as sparsity or group sparsity, to further improve recovery guarantees and efficiency. For instance, the sparse plus low-rank ($S$ + $L$) approach proposed by 
\cite{lly+12}, and its variants, such as the robust principal component analysis (RPCA) and the sparse subspace clustering (SSC), provide efficient solutions with improved robustness to outliers and noise. More recently, \cite{pkcs17} considers the non-square matrix sensing under RIP assumptions, and show that matrix factorization 
does not introduce any spurious
local minima 
under RIP. \cite{wr21} studies the technique of discrete-time mirror descent utilized to address the unregularized empirical risk in matrix sensing.

\paragraph{Compressive Sensing}
 
Compressive sensing has been a widely studied topic in signal processing and theoretical computer science field \cite{hikp12a, hikp12b, ik14, ps15, k16, ckps16, k17, ns19, nsw19,akm+19}. \cite{hikp12a} gave a fast algorithm (runs in time $O(k\log n \log( n/k))$ for generall in puts and $O(k\log n \log(n/k))$ for at most $k$ non-zero Fourier coefficients input) for $k$-sparse approximation to the discrete Fourier transform of an $n$-dimensional signal. \cite{k16} provided an algorithm such that it uses $O_d(k \log N \log\log N)$ samples of signal and runs in time $O_d(k\log^{d+3} N)$ for $k$-sparse approximation to the Fourier transform of a length of $N$ signal. Later work \cite{k17} proposed a new technique for analysing noisy hashing schemes that arise in Sparse FFT, which is called isolation on average, and applying it, it achieves sample-optimal results in $k\log^{O(1)}n$ time for estimating the values of a list of frequencies using few samples and computing Sparse FFT itself. 
\cite{ns19} gave the first sublinear-time $\ell_2/\ell_2$ compressed sensing which achieves the optimal number of measurements without iterating. After that, \cite{nsw19} provided an algorithm which uses $O(k \log k \log n) $ samples to compute a $k$-sparse approximation to the $d$-dimensional Fourier transform of a length $n$ signal. 
Later by \cite{sswz22a} provided an efficient Fourier Interpolation algorithm that improves the previous best algorithm \cite{ckps16} on sample complexity, time complexity and output sparsity. And in \cite{sswz22b} they presented a unified framework for the problem of band-limited signal reconstruction and achieves high-dimensional Fourier sparse recovery and high-accuracy Fourier interpolation. 
Recent work \cite{jls23} designed robust algorithms for super-resolution imaging that are efficient in terms of both running time and sample complexity for any constant dimension under the same noise model as \cite{ps15}, based on new techniques in Sparse Fourier transform.

\paragraph{Faster Iterative Algorithm via Sketching}

Low rank matrix completion is a well-known problem in machine learning with various applications in practical fields such as recommender systems, computer vision, and signal processing. Some notable surveys of this problem are provided in \cite{k09, nks19}. While Candes and Recht \cite{cr12} first proved the sample complexity for low rank matrix completion, other works such as \cite{ct10} and \cite{jns13} have provided improvements and guarantees on convergence for heuristics. In recent years, sketching has been applied to various machine learning problems such as linear regression \cite{cw13, nn13}, low-rank approximation \cite{cw13, nn13}, weighted low rank approximation, matrix CUR decomposition \cite{bw14, swz17, swz19}, and tensor regression \cite{dssw18, djs+19, swyz21, rsz22}, leading to improved efficiency of optimization algorithms in many problems. For examples, linear programming \cite{cls19,sy21,jswz21,dly21,gs22}, matrix completion \cite{gsyz23}, empirical risk minimization \cite{lsz19,qszz23}, training over-parameterized neural network \cite{bpsw21,szz21,jms+22,z22}, discrepancy algorithm \cite{z22,sxz22,dsw22}, frank-wolfe method \cite{xss21,sxyz22}, and reinforcement learning \cite{ssx21}.

\paragraph{Roadmap.}
We organize the following paper as follows. In Section~\ref{sec:tech_ov} we provide the technique overview for our paper. In Section~\ref{sec:prel} we provide some tools and existing results for our work. In Section~\ref{sec:ana} we provide the detailed analysis for our algorithm. In Section~\ref{sec:meas} we argue that our measurements are good. In Section~\ref{sec:shrink} we provide analysis for a shrinking step. 
In Section~\ref{sec:mat_sens_reg} we provide the analysis for our techniques used to solve the optimization problem at each iteration.

%% file: tech.tex
\section{Technique Overview}
\label{sec:tech_ov}

In this section, we provide a detailed overview of the techniques used to prove our results. Our approach is based on a combination of matrix sketching and low-rank matrix recovery techniques. Specifically, we use a sketching technique that allows us to efficiently extract relevant information from linear measurements of the low-rank matrix. We then use this information to recover the low-rank matrix using a convex optimization algorithm. With these techniques, we are able to improve previous results in both sample complexity and running time. From the two perspective, we give the overview of our techniques here. 

\subsection{Tighter Analysis Implies Reduction to Sample Complexity}
Our approach achieves this improvement by using a new sketching technique that compresses the original matrix into a smaller one while preserving its low-rank structure. This compressed version can then be used to efficiently extract relevant information from linear measurements of the original matrix.

To analyze the performance of our approach, we use tools from random matrix theory and concentration inequalities. Specifically, we use the Bernstein's inequality for matrices to establish bounds on the error of our recovery algorithm. 
We first define our measurements and operators, for each $i \in [m]$, let $x_i, y_i$ denotes samples from $\N(0, I_d)$. We define
\begin{itemize}
    \item $A_i := x_i y_i^\top$;
    \item $b_i := x_i^\top W_* y_i$;
    \item $W_0 := \frac{1}{m}\sum_{i = 1}^m b_i A_i$;
    \item $B_x := \frac{1}{m}\sum_{i = 1}^m (y_i^\top v)^2 x_ix_i^\top$;
    \item $B_y := \frac{1}{m}\sum_{i = 1}^m (x_i^\top v)^2 y_iy_i^\top$;
    \item $G_x := \frac{1}{m}\sum_{i=1}^m (y_i^\top v)(y_i^\top v_{\bot})x_ix_i^\top$;
    \item $G_x := \frac{1}{m}\sum_{i=1}^m (x_i^\top v)(x_i^\top v_{\bot})y_iy_i^\top$.
\end{itemize}
We need to argue that our measurements are \emph{good} under our choices of $m$, here the word ``good'' means that 
\begin{itemize}
    \item $\|W_0 - W_*\| \le \epsilon \cdot \|W_*\|$;
    \item $\|B_x- I\|\le \epsilon$ and $\|B_y - I\| \le \epsilon$;
    \item $\|G_x\| \le \epsilon$ and $\|G_y\| \le \epsilon$.
\end{itemize}
In our analysis we need to first bound $\|Z_i\|$ and $\|\E[Z_iZ_i^\top]\|$, where $Z_i := x_ix_i^\top U_*\Sigma_*Y_*^\top y_iy_i^\top$. With an analysis, we are able to show that (Lemma~\ref{lem:upper_bound_norm_of_Z_i} and Lemma~\ref{lem:norm_of_E_Z_i_Z_i_top})
\begin{align*}
    \Pr[\|Z_i\| \le C^2 k^2 \log^2(d/\delta)\sigma^4 \cdot \sigma_1^*] &~ \ge 1 - \delta/\poly(d) \\
    \|\E[Z_iZ_i^\top]\| &~ \le C^2k^2\sigma^4(\sigma_1^*)^2.  
\end{align*}
Now, applying these two results and by Bernstein's inequality, we are able to show that our operators are all ``good'' (Theorem~\ref{thm:main_measurement}).

\subsection{Induction Implies Correctness}

To get the final error bounded, we use an inductive strategy to analyze. Here we let $U_*$ and $V_*$ be the decomposition of ground truth $W_*$, i.e., $W_* = U_* V_*$. We show that, when iteratively applying our alternating minimization method, if $U_t$ and $V_t$ are closed to $U_*$ and $V_*$ respectively, then the output of next iteration $t+1$ is close to $U_*$ and $V_*$. Specifically, we show that, if $\dist(U_t, U_*) \le \frac{1}{4}\cdot \dist(V_t, V_*)$, then it yields
\begin{align}
    \dist(V_{t+1}, V_*) \le \frac{1}{4}\cdot\dist(U_t, U_*). 
\end{align}
Similarly, from the other side, if $\dist(V_{t+1}, V_*) \le \frac{1}{4} \cdot \dist(U_t, U_*)$, we have
\begin{align}
\label{eq:boud_U_tech_ov}
    \dist(U_{t+1}, U_*) \le \frac{1}{4} \cdot \dist(V_{t+1}, V_*).
\end{align}
This two recurrence relations together give the guarantee that, if the starting error $U_0 - U_*$ and $V_0 - V_*$, the distance from $V_t$ and $U_t$ to $V_*$ and $U_*$, respectively.

To prove the result, we first define the value of $\epsilon_d$ as $/10$. Then, by the algorithm, we have the following relationship between $V_{t+1}$ and $\wh{V}_{t+1} R^{-1}$,
\begin{align*}
    V_{t+1} = \wh{V}_{t+1} R^{-1} = (W_*^\top U_t - F)R^{-1},
\end{align*}
where the second step follows from the definition of $\wh{V}$ and defining $F$ as Definition~\ref{def:B_C_D_S}. Now we show that, $\|F\|$ and $\|R^{-1}\|$ can be bound respectively,
\begin{align*}
    \|F\| & ~ \le 2\epsilon k^{1.5} \cdot \sigma_1^* \cdot \dist(U_t, U_*)& \text{Lemma~\ref{lem:upper_bound_norm_F}}\\
    \|R^{-1}\| & ~ \le 10/\sigma_{k^*} &~\text{Lemma~\ref{lem:upper_bound_R_inverse_norm}}
\end{align*}
Note that the bound of $\R^{-1}$ need $\dist(U_t, U_*) \le \frac{1}{4}\cdot \dist(V_t, V_*))$

With these bounds, we are able to show the bound for $\dist(V_{t+1}, V_*)$. We first notice that, $\dist(V_{t+1}, V_*)$ can be represented as $(V_{*,\bot})^\top V_{t+1}$, where $V_{*,\bot} \in \R^{d \times (d-k)}$ is a fixed orthonormal basis of the subspace orthogonal to $\mathrm{span}(V_*)$. Then we show that (Claim~\ref{cla:V_*_bot_dot_V_t+1})
\begin{align*}
    (V_{*,\bot})^\top V_{t+1} = -(V_{*, \bot})^\top FR^{-1}. 
\end{align*}
Now, by turning $\dist(V_{t+1}, V_*)$ to the term of $F$ and $R$, and using the bound for $\|F\|$ and $\|R^{-1}\|$, we are finally able to reach the bound 
\begin{align*}
        \dist(V_{t+1}, V_*) 
    = &~ \|FR^{-1}\| \\
    \le &~ \|F\| \cdot \|R^{-1}\| \\
    \le &~ 2\epsilon k^{1.5} \cdot \sigma_1^* \cdot \dist(U_t, U_*) \cdot \|R^{-1}\| \\
    \le &~ 2\epsilon k^{1.5} \cdot \sigma_1^* \cdot \dist(U_t, U_*) \cdot 10/\sigma_k^* \\
    \le &~ 0.01 \cdot \dist(U_t, U_*).
\end{align*}
By a similar analysis, we can show Eq.\eqref{eq:boud_U_tech_ov}. 

Now applying them and with a detailed analysis, we have the claimed proved. Finally, when we prove that the initialization of the parameters are good, we can show that, the final output $W_T$ satisfies
\begin{align*}
    \|W_T - W_*\| \le \epsilon_0. 
\end{align*}

\subsection{Speeding up with Sketching Technique}
Now we consider the running time at each iteration. 
At each iteration of our algorithm, we need to solve the following optimization problem: 
\begin{align}
\label{eq:opt_in_it}
    \arg\min_{V \in \R^{d \times k}} \sum_{i = 1}^m (\tr[A_i^\top UV^\top] - b)^2.
\end{align}
When this problem is straightforwardly solved, it costs $O(md^2k^2)$ time, which is very expensive. So from another new direction, we give an analysis such that, this problem can be converted to a minimization problem where the target variable is a vector. To be specific, we show that, above optimization question \eqref{eq:opt_in_it} is equivalent to the following (Lemma~\ref{lem:equiv_regression}),
\begin{align*}
    \arg\min_{v \in \R^{dk}} \|Mv - b\|_2^2,
\end{align*}
where the matrix $M \in \R^{m \times dk}$
is defined to be the reformed matrix of $U^\top A_i$'s, i.e.,
\begin{align*}
    M_{i,*} := \vect(U^\top A_i), ~~\forall i \in [m].
\end{align*}
When working on this form of optimization problem, inspired by a recent work \cite{gsyz23}, we apply the fast sketch-to-solve low-rank matrix completion method. With this technique, we are able to reduce the running time to $\wt{O}(md^2k)$ (Theorem~\ref{thm:main_cost}), which is much more acceptable.

%% file: preli.tex
\section{Preliminary}
\label{sec:prel}
In this section, we provide preliminaries to be used in our paper.  In Section~\ref{sec:notation} we introduce notations we use. In Section~\ref{sec:rand_fac} and Section~\ref{sec:alg_fac} we provide some randomness facts and algebra facts respectively. In Section~\ref{sec:rip} we introduce the important definition of restricted isometry property. In Section~\ref{sec:rank_1_est} we provide results fro rank-one estimation. In Section~\ref{sec:rank1_ig_op} we introduce the rank-one independent Gaussian operator. In Section~\ref{def:angle_and_distance} we state our notations for angles and distances. In Section~\ref{sec:mat_conc} we provide some matrix concentration results. 

\subsection{Notations}
\label{sec:notation}

Let $x \in \R^n$ and $w \in \R_{\geq 0}^n $, we define the norm $\| x\|_w := (\sum_{i=1}^n w_i x_i^2)^{1/2}$.  

For $n > k$, for any matrix $A \in \R^{n \times k}$, we denote the spectral norm of $A$ by $\| A \|$, i.e., $\| A \| := \sup_{x\in\R^k} \| A x \|_2 / \| x \|_2$.

We denote the Frobenius norm of $A$ by $\| A \|_F$, i.e., $\| A \|_F : = (\sum_{i=1}^n \sum_{j=1}^k A_{i,j}^2 )^{1/2}$.

For any square matrix $A \in \R^{n \times n}$, we denote its trace by $\tr[A]$, i.e., $\tr[A] := \sum_{i=1}^n A_{i,i}$.

For any $A \in \R^{n \times d}$ and $B \in \R^{n \times d}$, we denote $\langle A , B \rangle = \tr[A^\top B]$.

Let $A \in \R^{n \times d}$ and $x \in \R^d$ be any matrix and vector, we have that
\begin{align*}
\| A x \|_2^2 = \langle A x, A x \rangle = \langle x , A^\top A x \rangle = x^\top A^\top A x.
\end{align*}

Let the SVD of $A \in \R^{n \times k}$ to be $U\Sigma B^\top$, where $U \in \R^{n \times k}$ and $V \in \R^{k \times k}$ have orthonormal columns and $\Sigma \in \R^{k \times k}$ be diagonal matrix. We say the columns of $U$ are the singular vectors of $A$. We denote the Moore-Penrose pseudoinverse matrix of $A$ as $A^\dagger \in \R{k \times n}$, i.e., $A^\dagger := V\Sigma^{-1}U^\top$. We call the diagonal entries $\sigma_1, \sigma_2, \dots, \sigma_k$ of $\Sigma$ to be the eigenvalues of $A$. We assume they are sorted from largest to lowest, so $\sigma_i$ denotes its $i$-th largest eigenvalue, and we can write it as $\sigma_i(A)$.

For $A \in \R^{n_1 \times d_1}$, $B \in \R^{n_2 \times d_2}$. We define kronecker product $\otimes$ as $(A \otimes B)_{i_1+(i_2-1)n_1, j_1 + (j_2-1)n_2}$
 
for all $i_1 \in [n_1]$, $j_1 \in [d_1]$, $i_2 \in [n_2]$ and $j_2 \in [d_2]$.

For any non-singular matrix $A \in \R^{n \times n}$, we define $A=QR$ its QR-decomposition, where $Q \in \R^{n \times n}$ is an orthogonal matrix and $R \in \R^{n \times n}$ is an non-singular lower triangular matrix. For any full-rank matrix $A \in \R^{n \times m}$, we define $A=QR$ its QR-decomposition, where $Q \in \R^{m \times n}$ is an orthogonal matrix and $R \in \R^{n \times n}$ is an non-singular lower triangular matrix. We use $R=$QR$(A) \in \R^{n \times n}$ to denote the lower triangular matrix obtained by the QR-decomposition of $A \in \R^{m \times n}$. 

Let $A \in \R^{k\times k}$ be a symmetric matrix. The eigenvalue decomposition of $A$ is $A = U\Lambda U^\top$, where $\Lambda$ is a diagonal matrix.

If a matrix $A$ is positive semidefinite (PSD) matrix, we denote it as $A \succeq 0$, which means $x^\top A x \ge 0$ for all $x$. 

Similarly, we say $A \succeq B$ if $x^\top  Ax \geq x^\top B x$ for all vector $x$.

For any matrix $U \in \R^{n \times k}$, we say $U$ is an orthonormal basis if $\| U_{i} \|=1$ for all $i \in [k]$ and for any $i\neq j$, we have $\langle U_i, U_j \rangle = 0$. Here for each $i \in [k]$, we use $U_i$ to denote the $i$-th column of matrix $U$.

For any $U \in \R^{n \times k}$ (suppose $n > k$)which is an orthonormal basis, 
we define $U_{\bot} \in \R^{n \times (n-k)}$ to be another orthonormial basis that, 
\begin{align*}
    U U^\top + U_{\bot} U_{\bot}^\top = I_n
\end{align*}
and
\begin{align*}
    U^\top U_{\bot} = {\bf 0}^{k \times (n-k)}
\end{align*}
where we use ${\bf 0}^{k \times (n-k)}$ to denote a $k \times (n-k)$ all-zero matrix. 

We say a vector $x$ lies in the span of $U$, if there exists a vector $y$ such that $x = U y$.

We say a vector $z$ lies in the complement of span of $U$, if there exists a vector $w$ such that $z = U_{\bot} w$. Then it is obvious that $\langle x,z \rangle = x^\top z =z^\top x =0$.

For a matrix $A$, we define $\sigma_{\min}(A) := \min_{x} \| A x \|_2 / \| x \|_2$. Equivalently,  $\sigma_{\min}(A) := \min_{x: \| x \|_2=1} \| A x \|_2$.

Similarly, we define $\sigma_{\max}(A) := \max_{x } \| A x \|_2 / \| x \|_2$. Equivalently,  $\sigma_{\max}(A) := \max_{x: \| x \|_2=1} \| A x \|_2$ 

Let $A_1, \cdots, A_n$ denote a list of square matrices. Let $S$ denote a block diagonal matrix $S = \begin{bmatrix} A_1 & & & \\
& A_2 & & \\
& & \ddots & \\
& & & A_n
\end{bmatrix}$. Then $\| S \| = \max_{i\in [n]} \| A_i \|$. 

We use $\Pr[]$ to denote probability. We use $\E[]$ to denote expectation.

Let $a$ and $b$ denote two random variables. Let $f(a)$ denote some event that depends on $a$ (for example $f(a)$ can be $a=0$ or $a \geq 10$.). Let $g(b)$ denote some event that depends on $b$. We say $a$ and $b$ are independent if $\Pr[f(a) \text{~and~} g(b)] = \Pr[f(a)] \cdot \Pr[g(b)]$. We say $a$ and $b$ are not independent if $\Pr[ f(a) \text{~and~} g(b)] \neq \Pr[f(a)] \cdot \Pr[g(b)]$. Usually if $a$ and $b$ are independent, then we also have $\E[ab] = \E[a] \cdot \E[b]$. 

We say a random variable $x$ is symmetric if $\Pr[x = u] = \Pr[x=-u]$.

For any random variable $x \sim {\cal N}(\mu,\sigma^2)$. This means $\E[x ] = \mu$ and $\E[x^2] = \sigma^2$.

We use $\wt{O}(f)$ to denote $f \cdot \poly(\log f)$.

\begin{definition}\label{def:Tmat}
We use $\Tmat(a,b,c)$ to denote the time of multiplying an $a \times b$ matrix with another $b \times c$ matrix.
 
\end{definition}  
We use $\omega$ to denote the exponent of matrix multiplication, i.e., $n^{\omega} =\Tmat(n,n,n)$.

\subsection{Randomness Facts}
\label{sec:rand_fac}

\begin{fact}\label{fac:random}
We have
\begin{itemize}
    \item Part 1. Expectation has linearity, i.e., $\E[ \sum_{i=1}^n x_i ] = \sum_{i=1}^n \E[x_i]$.
    \item Part 2. For any random vectors $x$ and $y$, if $x$ and $y$ are independent, then for any fixed function $f$, we have $\E_{x,y}[f(x) f(y)] = \E_x[f(x) ] \cdot \E_y[ f(y)]$.
    \item Part 3. Let $A\in \R^{d \times d}$ denote a fixed matrix. For any fixed function $f : \R^d \rightarrow \R^{d \times d}$, we have $\E_x[f(x) \cdot A ] = \E_x [f(x)] \cdot A$.
    \item Part 4. Given $n$ events $A_1, A_2, \cdots A_n$. For each $i \in [n]$, if $\Pr[ A_i ] \geq 1-\delta_i$. Then taking a union bound over all the $n$ events, we have $\Pr[ A_1 \text{~and~} A_2 \cdots A_n] \geq 1- \sum_{i=1}^n \delta_i$.
\end{itemize}
\end{fact}

\subsection{Algebra Facts}
\label{sec:alg_fac}
We state some standard facts and omit their proofs, since they're very standard.

\begin{fact}\label{fac:norm}
We have

\begin{itemize}
    \item For any orthonormal basis $U \in \R^{n \times k}$, we have $\| U x \|_2 = \| x \|_2$.
    \item For any orthonornal basis $U \in \R^{n \times k}$, we have $\| U \|_F \leq \sqrt{k}$.
    \item For any diagonal matrix $\Sigma \in \R^{k \times k}$ and any vector $x \in \R^k$, we have $\| \Sigma x \|_2 \geq \sigma_{\min}(\Sigma) \| x \|_2$. 
    \item For symmetric matrix $A$, we have $\sigma_{\min}(A) = \min_{z : \| z \|_2=1} z^\top A z$.
    \item For symmetric matrix $A$, we have $\sigma_{\min}(A) \| z\|_2^2 \leq z^\top A z$ for all vectors $z$.
    \item For symmetric matrix $A$, we have $\sigma_{\max}(A) \| z\|_2^2 \geq z^\top A z$ for all vectors $z$.
    \item For any matrix $A$, we have $\|A\| \leq \| A \|_F$.
    \item For any square matrix $A \in \R^{k \times k}$ and vector $x \in \R^{k}$, we have $x^\top A x = \sum_{i=1}^k \sum_{j=1}^k x_i A_{i,j} x_j = \sum_{i=1}^k x_i A_{i,i} x_i + \sum_{i\neq j} x_i A_{i,j} x_j$.
    \item For any square and invertible matrix $R$, we have $\| R^{-1} \| = \sigma_{\min}(R)^{-1}$
    \item For any matrix $A$ and for any unit vector $x$, we have $\| A \| \geq \| A x \|_2$.
    \item For any matrix $A$, $\| A A^\top \| = \| A^\top A \|$.
\end{itemize}
\end{fact}

\subsection{Restricted Isometry Property}
\label{sec:rip}

\begin{definition}[Restricted isometry property (RIP), see Definition 1 in \cite{zjd15}]
    A linear operator $\mathcal{A}: \R^{d\times d}\to\R^m$ satisfies $\mathrm{RIP}$ iff, for $\forall W \in \R^{d \times d}$  
    s.t. $\rank(W)\leq k$, the following holds:
    \begin{align*}
        (1-\epsilon_k) \cdot \|W\|_F^2\leq\|{\cal A}(W)\|_F^2\leq(1+\epsilon_k) \cdot \|W\|_F^2
    \end{align*}
    where $\epsilon_k > 0$ is a constant dependent only on $k$.
\end{definition}

\subsection{Rank-one Estimation}
\label{sec:rank_1_est}
The goal of matrix sensing is to design a linear operator $\mathcal{A}:\R^{d \times d} \to \R^m$ and a recovery algorithm so that a low-rank matrix $W_{*} \in \R^{d \times d}$ can be recovered exactly using $\mathcal{A}(W_*)$. 

\begin{definition}[Low-rank matrix estimation using rank one measurements]\label{def:b_i}
Given a ground-truth matrix $W_* \in \R^{d \times d}$. Let $(x_1, y_1) , \cdots,  (x_m, y_m) \in \R^{d} \times \R^d$ denote $m$ pair of feature vectors. Let $b \in \R^m$ be defined
\begin{align*}
b_i = x_i^\top W_* y_i, ~~~ \forall i \in [m].
\end{align*}
The goal is to use $b \in \R^m$ and $\{ (x_i,y_i)\}_{i \in [m]} \subset \R^d \times \R^d$ to recover $W_* \in \R^{d \times d}$.
\end{definition}

We propose two different kinds of rank-one measurement operators based on Gaussian distribution.

\subsection{Rank-one Independent Gaussian Operator}
\label{sec:rank1_ig_op}

We formally define Gaussian independent operator, here.
\begin{definition}[Gaussian Independent (GI) Operator]\label{def:gaussian_independent_operator}
Let $(x_1, y_1) , \cdots, (x_m, y_m) \subset \R^d \times \R^d$ denote i.i.d. samples from  Gaussian distribution.

For each $i \in [m]$, we define $A_i \in \R^{d \times d}$ as follows
\begin{align*}
A_i := x_i y_i^\top .
\end{align*} 

We define ${\cal A}_{\mathrm{GI}} \in \R^{d \times m d}$ as follows: 
\begin{align*}
\mathcal{A}_{\mathrm{GI}} := \begin{bmatrix} A_1 & A_2 & \cdots & A_m 
\end{bmatrix} .
\end{align*}
Here $\mathrm{GI}$ denotes Gaussian Independent.

\end{definition}

\subsection{Matrix Angle and Distance}
\label{sec:ang_dist}

We list several basic definitions and tools in literature, e.g., see \cite{gsyz23}.
\begin{definition}[Definition~4.1 in \cite{gsyz23}]\label{def:angle_and_distance}
Let $X, Y \in \R^{n \times k}$ denote two matrices.

For any matrix $X$, and for orthonormal matrix $Y$ ($Y^\top Y = I_k$) we define
\begin{itemize}
    \item $\tan \theta(Y,X) := \| Y_{\bot}^\top X ( Y^\top X )^{-1} \|$
\end{itemize}
For orthonormal matrices $Y$ and $X$ ($Y^\top Y = I_k$ and $X^\top X = I_k$), we define
\begin{itemize}
    \item $\cos \theta (Y,X) := \sigma_{\min} (Y^\top X)$. 
    \begin{itemize} 
        \item It is obvious that $\cos (Y,X) = 1/ \| (Y^\top X)^{-1} \|$ and $\cos(Y,X) \leq 1$.
    \end{itemize}
    \item $\sin \theta(Y,X) := \| (I - Y Y^\top) X \|$.
    \begin{itemize} 
        \item It is obvious that $\sin \theta(Y,X) = \| Y_{\bot} Y_{\bot}^\top X \| = \| Y_{\bot}^\top X \|$ and $\sin \theta(Y,X) \leq 1$.
        \item From Lemma~\ref{lem:sin^2_and_cos^2_is_1}, we know that $\sin^2\theta(Y,X) + \cos^2\theta(Y,X) = 1$. 
    \end{itemize}
    \item $\dist(Y,X) := \sin\theta(Y,X)$
\end{itemize}  

\end{definition}

\begin{lemma}[Lemma~A.7 in \cite{gsyz23}]\label{lem:tan_is_sin_cos}
Let $X, Y\in \R^{n\times k}$ be orthogonal matrices, then 
\begin{align*}
    \tan \theta(Y,X) = \frac{\sin \theta(Y,X)}{\cos \theta(Y,X)}.
\end{align*}
\end{lemma}

\begin{lemma}[Lemma~A.8 in \cite{gsyz23}]\label{lem:sin^2_and_cos^2_is_1}
Let $X, Y\in \R^{n\times k}$ be orthogonal matrices, then 
\begin{align*}
\sin^2\theta(Y, X) + \cos^2\theta(Y,X) =1.
\end{align*}
\end{lemma}

\subsection{Matrix Concentration}
\label{sec:mat_conc}

\begin{theorem}[Matrix Bernstein Inequality, Theorem 1.6 of \cite{t12}]\label{thm:matrix_bernstein_inequality}
  Given a finite sequence $\{ X_1, \cdots X_m \} \subset \R^{n_1 \times n_2}$ of independent, random matrices all with the dimension of $n_1 \times n_2$.

    Let $Z = \sum_{i=1}^m X_i$.

  Assume that
  \begin{align*}
    \E[X_i] = 0, \forall i \in [m], \| X_i \| \leq M, \forall i \in [m]
    \end{align*}

Let $\Var[Z]$ be the matrix variances statistic of sum
\begin{align*}
\Var[Z] = \max\{ \| \sum_{i=1}^m \E[X_iX_i^\top] \| , \| \sum_{i=1}^m \E[X_i^\top X_i] \| \}
\end{align*}
Then it holds that
\begin{align*}
\E[ \|Z \|] \leq (2 \Var[Z] \cdot \log(n_1+n_2))^{1/2} + M \log(n_1 + n_3) /3
\end{align*}
Further, for all $t>0$
\begin{align*}
\Pr[ \| Z \| \geq t ] \leq (n_1 + n_2) \cdot \exp( -\frac{t^2/2}{\Var[Z] + M t/3} )
\end{align*}
\end{theorem}

%% file: analysis.tex
\section{Analysis}
\label{sec:ana}
Here in this section, we provide analysis for our proposed algorithm. In Section~\ref{sec:ana_def}, we provide definitions in our algorithm analysis. In Section~\ref{sec:ana_op} we define the operators to be used. In Section~\ref{sec:ana_main} we provide our main theorem together with its proof. In Section~\ref{sec:main_indc_hypo} we introduce our main induction hypothesis. 

\subsection{Definitions}
\label{sec:ana_def}

\begin{algorithm}
    \caption{Our Faster Matrix Sensing Algorithm}\label{alg:main}
    \begin{algorithmic}[1]
    \Procedure{FastMatrixSensing}{${\cal A}_{all} \subset \R^{d \times d}, b_{all} \subset \R, \epsilon_0 \in (0,0.1),\epsilon \in (0,0.1), \delta \in (0,0.1)$} \Comment{Theorem~\ref{thm:main_informal}}
        \State   \Comment{Let $b_{all}$ scalar measurements}
        \State  \Comment{Let $\mathcal{A}_{all}$ sensing matrices}
        \State \Comment{Let $W_* \in \R^{d \times d}$ denote a rank-$k$ matrix}
        \State \Comment{Let $\sigma_1^*$ denote the largest singular value of $W_*$}
        \State \Comment{Let $\kappa$ denote the condition number of $W_*$}
        \State $T \gets \Theta( \log(k \kappa \sigma_1^*/\epsilon_0) )$
        \State $m \gets \Theta( \epsilon^{-2} (d+k^2) \log(d/\delta) )$
        \State Split $(\mathcal{A}_{all},b_{all})$ into $2T+1$ sets (each of size $m$) with $t$-th set being $\mathcal{A}^t \subset \R^{d \times d}$ and $b^t \in \R$
        \State  $U_0 \gets$  top-$k$ left singular vectors of $\frac{1}{m}\sum_{l=1}^mb_l^0 A_l^0$
        \For{$t=0$ to $T-1$}
            \State $b \gets b^{2t+1},\mathcal{A}\gets\mathcal{A}^{2t+1}$
            \State $\hat{V}_{t+1}\gets {\arg\min}_{V\in\R^{d\times k}}\sum_{l=1}^m(b_l-x_l^\top U_tV^\top y_l)^2$ \Comment{Using Lemma~\ref{lem:fast_solver}}
            \State $V_{t+1} \gets \mathrm{QR} (\hat{V}_{t+1})$\Comment{orthonormalization of $\hat{V}_{t+1}$}
            \State $b \gets b^{2t+2},\mathcal{A}\gets\mathcal{A}^{2t+2}$
            \State $\hat{U}_{t+1}\gets {\arg\min}_{U\in\R^{d\times k}}\sum_{l=1}^m(b_l-x_l^\top UV_{t+1}^\top y_l)^2$ \Comment{Using Lemma~\ref{lem:fast_solver}}
            \State $U_{t+1} \gets \mathrm{QR}(\hat{U}_{t+1})$\Comment{orthonormalization of $\hat{U}_{t+1}$}
        \EndFor
        \State  $W_T \gets U_T(\hat{V}_T)^\top$
        \State \Return $W_T$
    \EndProcedure
    \end{algorithmic}
\end{algorithm}

\begin{definition}\label{def:W_*}
We define $W_* \in \R^{d \times d}$ as follows
\begin{align*}
W_* = U_* \Sigma_* V_*^\top
\end{align*}
where $U_* \in \R^{n \times k}$ are orthonormal columns,  
and $V_* \in \R^{n \times k}$ are orthonormal columns.
Let $\sigma_1^*, \sigma_2^*, \cdots \sigma_k^*$ denote the diagonal entries of diagonal  matrix $\Sigma_* \in \R^{d \times d}$.
\end{definition}

\begin{definition}[Condition number]\label{def:kappa}
Let $W_*$ be defined as Definition~\ref{def:W_*}. We define $\kappa$ to the condition number of $W_*$, i.e.,
\begin{align*}
\kappa : = \sigma_1/\sigma_k.
\end{align*}
It is obvious that $\kappa \geq 1$.
\end{definition}

\begin{definition}[Measurements]\label{def:A_i_b_i}
For each $i \in [m]$, let $x_i,y_i$ denote samples from ${\cal N}(0,I_d)$.

For each $i \in [m]$, we define
\begin{align*}
A_i = x_i y_i^\top
\end{align*}
and
\begin{align*}
b_i = x_i^\top W_* y_i.
\end{align*}
\end{definition}

\subsection{Operators}
\label{sec:ana_op}

\begin{definition}[Initialization]\label{def:init}
For each $i \in [m]$, let $A_i$ and $b_i$ be defined as Definition~\ref{def:A_i_b_i}. 

 We define $W_0 := \frac{1}{m} \sum_{i=1}^m b_i A_i$.

We say initialization matrix $W_0 \in \R^{d \times d}$ is an $\epsilon$-good operator if 
\begin{align*}
    \| W_0 - W_* \| \leq \| W_* \| \cdot \epsilon.
\end{align*}

\end{definition}

\begin{definition}[Concentration of operators $B_x,B_y$]\label{def:B_operator} 
For any vectors $u,v$, we define  
\begin{itemize}
    \item $B_x:=\frac{1}{m}\sum_{l=1}^m(y_l^\top v)^2x_lx_l^\top$
    \item $B_y:=\frac{1}{m}\sum_{l=1}^m(x_l^\top u)^2 y_ly_l^\top$
\end{itemize}  
 We say $B = (B_x,B_y)$ is $\epsilon$-operator if the following holds: 

 \begin{itemize}
 \item $\|B_x-I\| \leq \epsilon$ 
 \item $\|B_y-I\| \leq \epsilon$
 \end{itemize}
\end{definition}

\begin{definition}[Concentration of operators $G_x,G_y$]\label{def:G_operator}
For any vectors $u,v \in \R^d$. 
We define  
\begin{itemize}
    \item $G_x:=\frac{1}{m}\sum_{l=1}^m(y_l^\top v)(y_l^\top v_\bot)x_lx_l^\top$ 
    \item $G_y:=\frac{1}{m}\sum_{l=1}^m(x_l^\top u)(x_l^\top u_\bot ) y_ly_l^\top$
\end{itemize}
 $u,u_\bot \in \R^{d},v,v_\bot\in \R^{d}$ are unit vectors, s.t., $u^\top u_\bot=0$ and $v^\top v_\bot=0$. 
 We say $G = (G_x,G_y) $ is $\epsilon$-operator if the following holds

 \begin{itemize}
 \item $\|G_x\|\leq\epsilon$,
 \item $\|G_y\|\leq\epsilon$.
 \end{itemize}
\end{definition}

\subsection{Main Result}
\label{sec:ana_main}

We prove our main convergence result as follows:  
\begin{theorem}[Formal version of Theorem~\ref{thm:main_informal}]\label{thm:main_convergence}
Let $W_* \in \R^{d \times d}$ be defined as Definition~\ref{def:W_*}.

Also, let $\mathcal{A}:\R^{d \times d} \to \R^m$ be a linear measurement operator parameterized by m matrices, i.e., $\mathcal{A}=\{A_1,A_2,\cdots,A_m\}$ where $A_l=x_l y_l^\top$. Let $\mathcal{A}(W)$ be as given by
\begin{align*}
    b=\mathcal{A}(W)=
    \begin{bmatrix}
        \tr[ A_1^\top W]& \tr[ A_2^\top W]  & \cdots & ~ \tr[A_m^\top W]
    \end{bmatrix}^\top
\end{align*}

If the following conditions hold 
\begin{itemize}
    \item $\epsilon= 0.001 / (k^{1.5} \kappa ) $ 
    \item $T =  100\log( \kappa k / \epsilon_0)$
    \item Let $\{(b_i,A_i)\}_{i\in [m]}$ be an $\epsilon$-init operator (Definition~\ref{def:init}).
    \item Let $B$ be an $\epsilon$-operator (Definition~\ref{def:B_operator}).  
    \item Let $G$ be an $\epsilon$-operator(Definition~\ref{def:G_operator}).
\end{itemize}
Then, after $T$-iterations of the alternating minimization method (Algorithm \ref{alg:main}), we obtain $W_T=U_T V_T^\top$ s.t., 
\begin{align*}
\|W_T-W_*\|\leq\epsilon_0.
\end{align*}
\end{theorem}
\begin{proof}

We first present the update equation for $\hat{V}_{t+1} \in \R^{d \times k}$.

Also, note that using the initialization property (first property mentioned in Theorem~\ref{thm:main_convergence}), we get, 
\begin{align*}
\|W_0 -W_*\|\leq \epsilon \sigma_1^* \leq \frac{\sigma_k^*}{100} .
\end{align*}

Now, using the standard sin theta theorem for singular vector perturbation \cite{l94}, we get: 
\begin{align*}
\dist(U_0,U_*) \leq & ~ \frac{1}{100} \\
\dist(V_0,V_*) \leq & ~ \frac{1}{100}
\end{align*}

After $T $ iteration (via Lemma~\ref{lem:main_induction}), we obtain
\begin{align*}
\dist(U_T,U_*) \leq & ~ (1/4)^T \\
\dist(V_T,V_*) \leq & ~ (1/4)^T
\end{align*}
which implies that
\begin{align*}
\| W_T - W_* \| \leq \epsilon_0
\end{align*}

\end{proof}

\subsection{Main Induction Hypothesis}
\label{sec:main_indc_hypo}

\begin{lemma}[Induction hypothesis]\label{lem:main_induction}
    We define $\epsilon_d: = 1/10$.
    We assume that $\epsilon= 0.001 / (k^{1.5} \kappa ) $.
    For all $t \in [T]$, we have the following results.
    
    \begin{itemize}
        \item Part 1. If $\dist(U_t,U_*) \leq \frac{1}{4} \dist(V_t, V_*) \leq  \epsilon_d$, then we have
        \begin{itemize}
            \item $\dist(V_{t+1}, V_*) \leq \frac{1}{4} \dist(U_t,U_*) \leq \epsilon_d$
        \end{itemize}
        \item Part 2. If $\dist(V_{t+1}, V_*) \leq \frac{1}{4} \dist(U_t,U_*) \leq \epsilon_d$, then we have
        \begin{itemize}
            \item $\dist(U_{t+1},U_*) \leq \frac{1}{4} \dist(V_{t+1}, V_*) \leq  \epsilon_d$
        \end{itemize}
    \end{itemize}
\end{lemma}
\begin{proof}

{\bf Proof of Part 1.}

Recall that for each $i \in [n]$, we have
\begin{align*}
b_i = x_i^\top W_{*} y_i = \langle x_i y_i^\top , W_* \rangle  = \langle A_i, W_* \rangle = \tr[A_i^\top W_*].
\end{align*}

Recall that  
\begin{align*}
\hat{V}_{t+1}=&~\arg\min_{V\in\R^{d\times k}}\sum_{i=1}^{m}(b_i-x_i^\top U_t V^\top y_i)^2\\
=&~\arg\min_{V\in\R^{d\times k}}\sum_{i=1}^{m}(x_i^\top W_* y_i-x_i^\top U_t V^\top y_i)^2
\end{align*}

Hence, by setting gradient of this objective function to zero. Let $F \in \R^{d \times k}$ be defined as Definition~\ref{def:B_C_D_S}.

We have $\wh{V}_{t+1} \in \R^{d \times k}$ can be written as follows: 
\begin{align}\label{eq:definition_of_hat_V}
    \hat{V}_{t+1} = W_*^\top U_t - F
\end{align}
where $F \in \R^{d \times k}$ is the error matrix
\begin{align*}
F = \begin{bmatrix} F_1 & F_2 &\cdots & F_k \end{bmatrix}
\end{align*}
where $F_i \in \R^d$ for each $i \in [k]$.

Then, using the definitions of $F \in \R^{d \times k}$ and Definition~\ref{def:B_C_D_S},  
we get:

\begin{align}
    \left[
    \begin{array}{cc}
         F_1  \\
         \vdots \\
         F_k
    \end{array}
    \right]
    =B^{-1}(BD-C)S\cdot \vect(V_*)
\end{align}
where $\vect(V_*) \in \R^{dk}$ is the vectorization of matrix $V_* \in \R^{d \times k}$.

Now, recall that in the $t+1$-th iteration of Algorithm~\ref{alg:main}, $V_{t+1} \in \R^{d \times k}$ is obtained by QR decomposition of $\hat{V}_{t+1} \in \R^{d \times k}$. Using notation mentioned above,
\begin{align}\label{eq:variation_of_hat_V_plus_one}
    \hat{V}_{t+1}=V_{t+1}R
\end{align}

where $R \in \R^{k \times k}$ denotes the lower triangular matrix $R_{t+1} \in \R^{k \times k}$ obtained by the QR decomposition of $V_{t+1} \in \R^{d \times k}$.

We can rewrite $V_{t+1} \in \R^{d \times k}$ as follows
\begin{align}\label{eq:variation_V}
    V_{t+1} = & ~ \hat{V}_{t+1} R^{-1} \notag \\
      = & ~ (W_*^\top U_t-F)R^{-1}
\end{align}
where the first step follows from  Eq.~\eqref{eq:variation_of_hat_V_plus_one} , and the last step follows from Eq.~\eqref{eq:definition_of_hat_V}.

Multiplying both the sides by $V_{*,\bot} \in \R^{d \times (d-k)}$, where $V_{*,\bot} \in \R^{d \times (d-k)}$ is a fixed orthonormal basis of the subspace orthogonal to $\mathrm{span}(V_*)$, using Claim~\ref{cla:V_*_bot_dot_V_t+1}
\begin{align}\label{eq:turn_V_t+1_into_F_R}
    (V_{*,\bot} )^\top V_{t+1} = -(V_{*,\bot} )^\top FR^{-1}
\end{align}

Thus, we get:
\begin{align*}
    \dist(V_{t+1}, V_*) =&~ \|(V_{*,\bot} )^\top V_{t+1}\| \notag \\
    = & ~ \| (V_{*,\bot} )^\top F R^{-1} \| \notag \\\
    = & ~ \| F R^{-1} \| \notag \\
    \leq & ~ \|F\| \cdot \|R^{-1}\| \\
    \leq & ~ 0.001 \sigma_k^* \dist(U_t, U_*) \cdot \| R^{-1} \| \\
    \leq & ~ 0.001 \sigma_k^* \dist(U_t, U_*) \cdot 2 (\sigma_k^*)^{-1} \\
    \leq & ~ 0.01 \cdot \dist(U_t,U_*)
\end{align*}

where the first step follows from definition of $\dist$ (see Definition~\ref{def:angle_and_distance}), the second step follows from Eq.~\eqref{eq:turn_V_t+1_into_F_R}, the third step follows from $V_{*,\bot}$ is an orthonormal basis, 
 and the forth step follows from  Fact~\ref{fac:norm}, the fifth step follows from Lemma.~\ref{lem:upper_bound_norm_F}, the sixth step follows from Lemma~\ref{lem:upper_bound_R_inverse_norm} (In order to run this lemma, we need to the condition of Part 1 statement to be holding), the last step follows from simple algebra.

{\bf Proof of Part 2.}

Similarly, we can prove this as Part 1.

\end{proof}

\section{Measurements are Good Operator}
\label{sec:meas}
In this section, we provide detailed analysis for our operators. First Section~\ref{sec:truncated_gaus} we introduce some standard results for truncated Gaussian. In Section~\ref{sec:bound_Zi} and Section~\ref{sec:bound_expect} we bound the term $\|Z_i\|$ and $\|\E[Z_iZ_i^\top]\|$ respectively. In Section~\ref{sec:good_meas_main} we state our main lemma. In Section~\ref{sec:good_init} we show that out initialization is good. In Section~\ref{sec:good_opB_G} we show our two operators are good.

\subsection{Tools for Gaussian}
\label{sec:truncated_gaus}

We state a standard tool from literature,
\begin{lemma}[Lemma 1 in \cite{lm00} 
]\label{lem:lm}
    Let $X \sim \mathcal{X}_k^2$ be a chi-squared distributed random variable with $k$ degrees of freedom. Each one has zero means and $\sigma^2$ variance. 
    
    Then it holds that
    \begin{align*}
        \Pr[X - k\sigma^2 \geq (2\sqrt{kt} + 2t) \sigma^2]
        \leq & ~ \exp{(-t)}\\
        \Pr[k\sigma^2 - X \geq 2\sqrt{kt}\sigma^2]
        \leq & ~ \exp{(-t)}
    \end{align*}
    Further if $k \geq \Omega(\epsilon^{-2} t)$ and $t \geq \Omega(\log(1/\delta))$, then we have
    \begin{align*}
    \Pr[ | X - k \sigma^2 | \leq \epsilon k \sigma^2 ] \leq \delta.
    \end{align*}
\end{lemma}

We state a standard fact for the 4-th moment of Gaussian distribution.
\begin{fact}
Let $x \sim {\cal N}(0,\sigma^2)$, then it holds that $\E_{x \sim {\cal N}(0,\sigma^2)}[x^4] = 3 \sigma^2$.
\end{fact}

\begin{lemma}\label{lem:gaussian_4th_vector_moment}
Let $x \sim {\cal N}(0, \sigma^2 I_d)$ denote a random Gaussian vector. Then we have
\begin{itemize}
\item Part 1 
\begin{align*}
\E[x x^\top x x^\top] = (d+2) \sigma^4
\end{align*}
\item Part 2
\begin{align*}
\| \E[ x x^\top x x^\top ]\| = (d+2) \sigma^4
\end{align*}
\end{itemize}
\end{lemma}
\begin{proof}
We define $A:=xx^\top xx^\top$. Then we have
\begin{align*}
A_{i,j} = x_i \sum_{l=1}^d x_l x_l x_j
\end{align*}
For $i=j$, we have
\begin{align*}
\E[A_{i,i}] = & ~ \E[ x_i \sum_{l=1}^d x_l x_l x_i ] \\
= & ~ \E[x_i(\sum_{l=1}^{i-1}x_l x_l + x_i x_i + \sum_{l=i+1}^d x_l x_l) x_i] \\
= & ~ \E[x_i^4] + \sum_{l \in [d] \backslash i} \E[x_l^2 x_i^2] \\
= & ~ \E[x_i^4] + \sum_{l \in [d] \backslash i} \E[x_l^2] \E[x_i^2] \\
= & ~ \E[x_i^4] + (d-1) \sigma^4 \\
= & ~ 3 \sigma^4 + (d-1) \sigma^4 \\
= & ~ (d + 2) \sigma^4
\end{align*}

where the third step follows from linearity of expectation (Fact~\ref{fac:random}), the forth step follows from $x_l$ and $x_i$ are independent, the fifth step follows $\E_{z \sim {\cal N}(0,\sigma^2)}[z^4] =3 \sigma^4$.

For $i\neq j$, we have
\begin{align*}
\E[A_{i,j}] = & ~ \E[ x_i \sum_{l=1}^d x_l x_l x_j ] \\
= & ~ \E[x_i x_j^3] + \E[x_i^3 x_j] + \sum_{l \in [d] \backslash i,j} \E[x_i x_l^2 x_j] \\
= & ~ 0
\end{align*}
where the second step follows from linearity of expectation (Fact~\ref{fac:random}).

\end{proof}

\begin{fact}[Rotation invariance property of Gaussian]
\label{fac:rot_inv_gaus}
    Let $A^\top \in \R^{d \times k}$ with $k < d$ denote an orthonormal basis (i.e., $AA^\top = I_k$). Then for a Gaussian $x \sim \N(0, \sigma^2 I_d)$, we have
    \begin{align*}
        Ax \sim \N(0, \sigma^2 I_k).
    \end{align*}
\end{fact}

\begin{proof}
    Let $y := Ax \in \R^k$, then
    \begin{align*}
        y_i = \sum_{j = 1}^dA_{ij}x_j, ~~\forall i \in [k]. 
    \end{align*}
    By definition of Gaussian distribution
 
    \begin{align*}
        y_i \sim \N(0, \sigma^2\sum_{j = 1}^dA_{ij}^2).
    \end{align*}
    Recall that $A^\top$ is an orthonormal basis.

    We have
    \begin{align*}
        A_{ij}^2 = 1.
    \end{align*}
    Thus we have
    \begin{align*}
        y \sim \N(0, \sigma^2 I_k),
    \end{align*}
\end{proof}

\subsection{Bounding \texorpdfstring{$\| Z_i \|$}{}}
\label{sec:bound_Zi}

\begin{lemma}\label{lem:upper_bound_norm_of_Z_i}
Let $x_i$ denote a random Gaussian vector samples from ${\cal N}(0, \sigma^2 I_d)$. Let $y_i$ denote a random Gaussian vector samples from ${\cal N}(0, \sigma^2 I_d)$.

Let $U_*, V_* \in \R^{d \times k}$.

We define
\begin{align*}
    Z_i := x_i x_i^\top U_* \Sigma_* V_*^\top y_i y_i^\top, ~~~\forall i \in [m]
\end{align*}
\begin{itemize}
\item Part 1. We have
\begin{align*}
    \Pr[ \| Z_i \| \leq C^2 k^2 \log^2(d/\delta) \sigma^4 \cdot \sigma_1^* ] \geq 1-\delta/\poly(d).
\end{align*}
\item Part 2. If $k \geq \Omega(\log(d/\delta))$  We have
\begin{align*}
    \Pr[ \| Z_i \| \leq C^2 k^2 \sigma^4 \cdot \sigma_1^* ] \geq 1-\delta/\poly(d).
\end{align*}
\end{itemize}
\end{lemma}
\begin{proof}

{\bf Proof of Part 1.}

We define 
\begin{align*}
a_i := U_*^\top x_i \in \R^k \\
b_i := V_*^\top y_i \in \R^k
\end{align*}
Since $U_*$ and $V_*$ are orthornormal basis, due to rotation invariance property of Gaussian (Fact~\ref{fac:rot_inv_gaus}) 
, we know that $a_i \sim {\cal N}(0,\sigma^2 I_k)$ and $b_i \sim {\cal N}(0, \sigma^2 I_k)$.

We also know that 

\begin{align*}
x_i = (U_*^\top)^{\dagger} a_i = U_* a_i \\
y_i = (V_*^\top)^{\dagger} b_i = V_* b_i
\end{align*}

Thus, by replacing $x_i,y_i$ with $a_i,b_i$, we have
\begin{align*}
\| Z_i \| 
= & ~ \| x_i x_i^\top U_* \Sigma_* V_*^\top y_i y_i^\top \| \\
= & ~ \| U_* a_i a_i^\top U_*^\top U_* \Sigma_* V_*^\top V_* b_i b_i^\top V_*^\top \| \\
= & ~ \| U_* a_i a_i^\top \Sigma_* b_i b_i^\top V_*^\top \| \\
\leq & ~ \| U_* \| \cdot \| a_i a_i^\top\| \cdot \| \Sigma_* \| \cdot \| b_i b_i^\top \| \cdot \| V_*^\top \| \\
\leq & ~ \sigma_1^* \cdot \| a_i \|_2^2 \cdot \| b_i \|_2^2
\end{align*}
where the second step follows from replacing $x,y$ by $a,b$, the third step follows from $U_*^\top U_* = I$ and $V_*^\top V_* = I$, the forth step follows from Fact~\ref{fac:norm}.

Due to property of Gaussian, we know that
\begin{align*}
\Pr[ |a_{i,j}| > \sqrt{C\log(d/\delta)} \sigma ] \leq \delta/\poly(d)
\end{align*}

Taking a union bound over $k$ coordinates, we know that
\begin{align*}
\Pr[ \| a_i \|_2^2 \leq C k \log(d/\delta) \sigma^2 ] \geq 1-\delta /\poly(d)
\end{align*}
Similarly, we can prove it for $\| b_i \|_2^2$.

{\bf Proof of Part 2.}
Since $k \geq \Omega(\log(d/\delta))$, then we can use Lemma~\ref{lem:lm} to obtain a better bound.

\end{proof}
\label{sec:bound_expect}

\subsection{Bounding \texorpdfstring{$\| \E[ Z_i Z_i^\top] \|$}{}}
\begin{lemma}\label{lem:norm_of_E_Z_i_Z_i_top}
We can show that
\begin{align*}
 \| \E[ Z_i Z_i^\top] \| \leq C^2 k^2 \sigma^4 (\sigma_1^*)^2.
\end{align*}
\end{lemma}

\begin{proof}

Using Lemma~\ref{lem:gaussian_4th_vector_moment}

\begin{align*}
\| \E_{a \sim {\cal N}(0, \sigma^2 I_k )}[ a_i a_i^\top a_i a_i^\top ] \| \leq C k \sigma^2.
\end{align*}
Thus, we have
\begin{align*} 
\E[ a_i a_i^\top a_i a_i^\top] \preceq Ck \sigma^2 \cdot I_k 
\end{align*}

Then, we have
\begin{align}\label{eq:bound_E_Z_i_Z_i_top}
         \| \E[{Z}_i Z_i^\top] \|
         = & ~ \| \E_{x,y}[ x_i x_i^\top U_* \Sigma_* V_*^\top y_i y_i^\top y_i y_i^\top V_* \Sigma_* U_*^\top x_i x_i^\top ] \| \notag \\
         = & ~ \|\E_{a,b}[ U_* a_i a_i^\top U_*^\top U_* \Sigma_* V_*^\top V_* b_i b_i^\top V_*^\top V_* b_i b_i^\top V_*^\top V_* \Sigma_* U_*^\top U_* a_i a_i^\top U_*^\top ]\| \notag \\
         = & ~ \| \E_{a,b}[ U_* a_i a_i^\top \Sigma_*  b_i b_i^\top  V_*^\top V_* b_i b_i^\top  \Sigma_*  a_i a_i^\top U_*^\top ] \| \notag \\ 
         = & ~ \| \E_{a,b}[ U_* a_i a_i^\top \Sigma_*  b_i b_i^\top b_i b_i^\top  \Sigma_*  a_i a_i^\top U_*^\top ] \| \notag \\ 
         \leq & ~ \|\E_{a,b}[ a_i a_i^\top \Sigma_*  b_i b_i^\top b_i b_i^\top  \Sigma_*  a_i a_i^\top  ]  \| \notag \\
         \leq & ~ \|\E_{a}[ a_i a_i^\top \Sigma_* \E_b[ b_i b_i^\top b_i b_i^\top ] \Sigma_*  a_i a_i^\top  ]  \| \notag \\
         \leq & ~ C^2 k^2 \sigma^4 (\sigma_1^*)^2
     \end{align}
     where the first step follows from the definition of $Z_i$, the second step follows from replacing $x_i,y_i$ with $a_i,b_i$, the third step follows from $U_*,V_*$ are orthonormal columns, the fourth step follows from $V_*$ are orthonormal columns, the fifth step follows from
    $\| U_* \| \leq 1$
    , the sixth step follows from
    using Lemma~\ref{lem:gaussian_4th_vector_moment} twice.
\end{proof}

\subsection{Main Results}
\label{sec:good_meas_main}

We prove our main result for measurements.  
\begin{theorem}[Formal of Theorem~\ref{thm:main_informal}, Measurements are good operator]\label{thm:main_measurement}

     Let $\{A_i,b_i\}_{i\in [m]}$ denote measurements be defined as Definition~\ref{def:A_i_b_i}.

     Assuming the following conditions are holding
     \begin{itemize}
        \item $k = \Omega(\log(d/\delta))$
        \item $m = \Omega(\epsilon^{-2} (d+k^2) \log(d/\delta))$
     \end{itemize} 
     
     Then, 
     \begin{itemize}
        \item The property in Definition~\ref{def:init}, initialization is a $\epsilon$-operator
        \item The property in Definition~\ref{def:B_operator}, $B$ are $\epsilon$-operator.
        \item The property in Definition~\ref{def:G_operator}, $G$ are $\epsilon$-operator.
     \end{itemize}
     holds with probability at least $1-\delta/\poly(d)$.

\end{theorem}
\begin{proof}
Using Lemma~\ref{lem:init_S_is_good} and Lemma~\ref{lem:B_G_operator_are_good}, we complete the proof.
\end{proof}

\subsection{Initialization Is a Good Operator}
\label{sec:good_init}
\begin{lemma}\label{lem:init_S_is_good}
We define matrix $S \in \R^{d \times d}$ as follows
\begin{align*}
    S: = \frac{1}{m} \sum_{i=1}^m b_i A_i.
\end{align*}

If the following two condition holds
 
\begin{itemize}
    \item Condition 1. $k = \Omega(\log(d/\delta))$,
    \item Condition 2. $m = \Omega( \epsilon^{-2} k^2 \log(d/\delta) )$.
\end{itemize}

Then we have
\begin{align*}
\Pr[ \| S  - W_* \| \leq \epsilon \cdot \| W_* \| ] \geq 1-\delta.
\end{align*}

\end{lemma}

\begin{proof}

    (Initialization in Definition~\ref{def:init}) Now, we have: 

    \begin{align*}
        S = & ~ \frac{1}{m} \sum_{i=1}^m b_i A_i \\
          = & ~ \frac{1}{m} \sum_{i=1}^m b_i x_i y_i^\top \\
          = & ~ \frac{1}{m} \sum_{i=1}^m  x_i b_i y_i^\top \\
          = & ~ \frac{1}{m} \sum_{i=1}^m x_i x_i^\top W_* y_i  y_i^\top \\ 
          = & ~ \frac{1}{m} \sum_{i=1}^m x_i x_i^\top U_* \Sigma_* V_*^\top y_i y_i^\top, \\
    \end{align*}
     where the first step follows from Definition~\ref{def:init}, the second step follows from $A_i = x_i y_i^\top$, the third step follows from $b_i$ is a scalar, the forth step follows from $b_i = x_i^\top W_* y_i$, the fifth step follows from $W_* = U_* \Sigma_* V_*^\top$. 

For each $i \in [m]$, we define matrix $Z_i \in \R^{d \times d}$ as follows:
\begin{align*}
Z_i := x_i x_i^\top U_* \Sigma_* V_*^\top y_i y_i^\top,
\end{align*}
then we can rewrite $S \in \R^{d \times d}$ in the following sense,
     \begin{align*}
         S = \frac{1}{m} \sum_{i=1}^m Z_i
     \end{align*}
     
     Note that, we can compute $\E[Z_i] \in \R^{d \times d}$
     \begin{align*}
        \E_{x_i,y_i}[Z_i]
        = & ~ \E_{x_i, y_i}[ \underbrace{ x_ix_i^\top }_{d \times d} \underbrace{ U_* \Sigma_* V_*^\top }_{ d \times d} \underbrace{ y_i y_i^\top }_{d \times d} ] \\
        = & ~ \E_{x_i}[ \underbrace{ x_ix_i^\top }_{d \times d} \underbrace{ U_* \Sigma_* V_*^\top }_{ d \times d} ]  \cdot \E_{y_i} [\underbrace{ y_i y_i^\top }_{d \times d} ] \\
        = & ~ \E_{x_i} [x_i x_i^\top ] \cdot U_* \Sigma_* V_*^\top \cdot \E_{y_i}[ y_i y_i^\top] \\ 
        = & ~ U_* \Sigma_* V_*^\top
    \end{align*}
    where the first step follows definition of $Z_i$, the second step follows from $x_i$ and $y_i$ are independent and Fact~\ref{fac:random}, the third step follows from Fact~\ref{fac:random} the forth step follows from $\E[x_ix_i^\top] = I_d$ and $\E[y_i y_i^\top] = I_d$.

     As $S \in \R^{d \times d}$ is a sum of $m$ random matrices, the goal is to apply Theorem~\ref{thm:matrix_bernstein_inequality}
    to show that $S$ is close to
    \begin{align*}
    \E[S] = & ~  W_* \\
    = & ~ U_* \Sigma_* V_*^\top
    \end{align*}
    for large enough $m$. 
    
Using Lemma~\ref{lem:upper_bound_norm_of_Z_i} (Part 2) with choosing Gaussian variance $\sigma^2=1$, we have
 
\begin{align}\label{eq:upper_bound_of_Z_i}
\Pr[ \| Z_i \| \leq C^2 k^2  \sigma_1^*, \forall i \in [m] ] \geq 1-\delta/\poly(d)
\end{align}

     Using Lemma~\ref{lem:norm_of_E_Z_i_Z_i_top} with choosing Gaussian variance $\sigma^2= 1$, we can bound $ \| \E[{Z}_i {Z}_i^\top] \|$ as follows  
     
     \begin{align}\label{eq:upper_bound_of_E_tilde_Z_i_Z_i_top}
         \| \E[{Z}_i Z_i^\top] \|
         \leq & ~ C^2 k^2 (\sigma_1^*)^2
     \end{align}

 Let $Z = \sum_{i=1}^m (Z_i - W_*)$.
 
 Applying Theorem~\ref{thm:matrix_bernstein_inequality} we get 
     \begin{align}\label{eq:pr_norm_Z_geq_t}
\Pr[ \| Z \| \geq t ] \leq 2d \cdot \exp( -\frac{t^2/2}{\Var[Z] + M t/3} )
\end{align}
     where

 \begin{align*}
 Z = & ~ m S - m W_* \\
 \Var[Z] = & ~ m \cdot C^2 k^2  (\sigma_1^*)^2, & \text{~by~Eq.~\eqref{eq:upper_bound_of_E_tilde_Z_i_Z_i_top}} \\
 M= & ~ C^2 k^2 \sigma_1^* & \text{~by~Eq.~\eqref{eq:upper_bound_of_Z_i}}
 \end{align*}

Replacing $t= \epsilon \sigma_1^* m$ and $Z = mS - mW_*$ inside $\Pr[]$ in Eq.~\eqref{eq:pr_norm_Z_geq_t}, we have 

\begin{align*}
\Pr[ \| S - W^* \| 
\geq & ~ \epsilon \sigma_1^* ] \leq 2d \cdot \exp\bigg( -\frac{t^2 /2}{\Var[Z] + M t /3} \bigg) 
\end{align*}
Our goal is to choose $m$ sufficiently large such that the above quantity is upper bounded by $2d \cdot \exp ( - \Omega( \log(d/\delta) ))$.

First, we need 

\begin{align*}
\frac{t^2}{\Var[Z]}
= & ~\frac { \epsilon^2 m^2 (\sigma_1^*)^2 }{ m \cdot C^2 k^2  (\sigma_1^*)^2  } \\
= & ~ \frac{ \epsilon^2 m }{ C^2 k^2  } \\
\geq & ~ \log(d/\delta)
\end{align*}
where the first step follows from choice of $t$ and bound for $\Var[Z]$.

This requires
\begin{align*}
m \geq C^2 \epsilon^{-2} k^2 \log(d/\delta)
\end{align*}

Second, we need 
\begin{align*}
\frac { t^2 }{  M t  } = & ~ \frac{ \epsilon m \sigma_1^* }{ M } \\
= & ~\frac{\epsilon m \sigma_1^* }{C^2 k^2   \sigma_1^*} \\
= & ~\frac{\epsilon m }{C^2 k^2  } \\
\geq & ~ \log(d/\delta)
\end{align*}
where the first step follows from choice of $t$ and the second step follows from bound on $M$.

This requires
\begin{align*}
m \geq C^2 \epsilon^{-2} k^2 \log(d/\delta)
\end{align*}
    
  Finally, we should choose
     \begin{align*}
         m \geq 10C^2 \epsilon^{-2} k^2 \log(d/\delta) ,  
     \end{align*}

    Which implies that 
     \begin{align}\label{eq:upper_bound_of_S}
         \Pr[ \| S - W_* \| \leq \epsilon \cdot \sigma_1^* ] \geq 1- \delta/\poly(d).
     \end{align}
     
   Taking the union bound with all $\| Z_i \|$ are upper bounded, then we complete the proof.

\end{proof}

\subsection{Operator \texorpdfstring{$B$}{} and \texorpdfstring{$G$}{} is good}
\label{sec:good_opB_G}

\begin{lemma}\label{lem:B_G_operator_are_good}
If the following two conditions hold
\begin{itemize}
    \item Condition 1. $d = \Omega(\log(d/\delta))$
    \item Condition 2. $m = \Omega(\epsilon^{-2} d \log(d/\delta))$
\end{itemize}
Then operator $B$ (see Definition~\ref{def:B_operator}) is $\epsilon$ good, i.e.,
\begin{align*}
\Pr[ \| B_x - I_d\| \leq \epsilon ] \geq & ~ 1-\delta/\poly(d)  \\
\Pr[ \| B_y - I_d \| \leq \epsilon ] \geq & ~ 1-\delta/\poly(d)  
\end{align*}
Similar results hold for operator $G$ (see Definition~\ref{def:G_operator}).
\end{lemma}

\begin{proof}

     Recall that
     $B_x:=\frac{1}{m}\sum_{l=1}^m(y_l^\top v)^2x_lx_l^\top$.

     Recall that $B_y:=\frac{1}{m}\sum_{l=1}^m(x_l^\top u)^2 y_ly_l^\top$.
     
     Now, as $x_i,y_i$ are rotationally invariant random variables 
     , wlog, we can assume $u=e_1$.
     
     We use $x_{i,1} \in \R$ to denote the first entry of $x_i \in \R^d$.

     Thus,  
     \begin{align*}
         (x_i^\top u u^\top x_i)=x_{i,1}^2
     \end{align*}
     Then 
     \begin{align*}
        \E[ (x_i^\top u u^\top x_i)^2 ] = \E[x_{i,1}^4 ] = 3
     \end{align*}

    We define
    \begin{align*}
    Z_i = (x_i^\top u)^2 y_i y_i^\top
    \end{align*}
    then
    \begin{align*}
        \E[Z_i] = I_d
    \end{align*}    
    
   Using similar idea in Lemma~\ref{lem:upper_bound_norm_of_Z_i}, we have
    \begin{align*}
    \Pr[ \| Z_i \| \leq C d , \forall i \in [m] ] \geq 1- \delta/\poly(d)
    \end{align*}
    
    We can bound
    \begin{align*}
    \| \E[ Z_i Z_i^\top ] \|
    = & ~ \| \E_{x,y}[ (x_i^\top u)^2 y_i y_i^\top  y_i y_i^\top  (x_i^\top u)^2 ] \| \\
    = & ~ \| \E_{x}[ (x_i^\top u)^2 \E_y[y_i y_i^\top  y_i y_i^\top ]  (x_i^\top u)^2 ] \| \\
    = & ~ (d+2) \cdot | \E_{x}[ (x_i^\top u)^2  (x_i^\top u)^2 ] | \\
    = & ~ (d+2) \cdot 3 \\
    \leq & ~ C d
    \end{align*}
where the fourth step follows from $C \geq 1$ is a sufficiently large constant.

Let $Z = \sum_{i=1}^m (Z_i - I_d)$.

    Applying Theorem~\ref{thm:matrix_bernstein_inequality} we get
    \begin{align*}
        \Pr[\| Z\| \ge t] \le 2d \cdot \exp(-\frac{t^2/2}{\Var[Z] + Mt/3}),
    \end{align*}
where
\begin{align*}
    Z = & ~ m \cdot B - m \cdot I \\
    \Var[Z] = & ~ C m d \\
    M = & ~ C d \\
\end{align*}

Using $t = m \epsilon$ and $Z = \sum_{i=1}^m (Z_i - I_d)$, and $B = \frac{1}{m} \sum_{i=1}^m Z_i$, we have
\begin{align*}
\Pr[ \| Z \| \geq t] 
= & ~ \Pr[ \| \sum_{i=1}^m (Z_i - I_d) \| \geq m \epsilon ] \\
= & ~ \Pr[ \| \frac{1}{m} \sum_{i=1}^m Z_i - I_d \| \geq \epsilon ] \\
= & ~ \Pr[ \| B - I_d \| \geq \epsilon ]
\end{align*}
    
By choosing $t = m \epsilon$ and $m = \Omega(\epsilon^{-2} d \log(d/\delta))$ we have

    \begin{align*}
        \Pr[ \| B - I_d \| \geq \epsilon ] \leq \delta/\poly(d).
    \end{align*}
where $B$ can be either $B_x$ or $B_y$.

    Similarly, we can prove 
    \begin{align*}
        \Pr[\|G_x\| \leq \epsilon] \geq 1 - \delta, \\ 
        \Pr[\|G_y\| \leq \epsilon] \geq 1 - \delta.
    \end{align*}

\end{proof}

%% file: shrink.tex
\section{One Shrinking Step}
\label{sec:shrink}
In this section, we provide a shirking step for our result. In Section~\ref{sec:def_BCDS} we define the matrices $B, C, D ,S$ to be used in analysis. In Section~\ref{sec:bdc_upper} we upper bound the norm of $BD- C$. In Section~\ref{sec:rewrite_V} we show the update term $V_{t + 1}$ can be written in a different way. In Section~\ref{sec:F_upper} and Section~\ref{sec:R_inv_upper} we upper bounded $\|F\|$ and $\|R^{-1}\|$ respectively. 

\subsection{Definitions of \texorpdfstring{$B,C,D,S$}{}}
\label{sec:def_BCDS}
\begin{definition}\label{def:B_C_D_S}

For each $p \in [k]$, let $u_{*,p} \in \R^n$ denotes the $p$-th column of matrix $U_* \in \R^{n \times k}$. 

For each $p \in [k]$, let $u_{t,p}$ denote the $p$-th column of matrix $U_t \in \R^{n \times k}$.

We define block matrices $B, C, D, S \in \R^{kd \times kd}$ as follows:
For each $(p,q) \in [k] \times [k]$
\begin{itemize}
    \item Let $B_{p,q} \in \R^{d \times d}$ denote the $(p,q)$-th block of $B$ 
    \begin{align*}
        B_{p,q}= \sum_{i=1}^m \underbrace{ y_i y_i^\top }_{d \times d \mathrm{~matrix}} \cdot \underbrace{ (x_i^\top u_{t,p})}_{ \mathrm{scalar} } \cdot \underbrace{ (x_i^\top u_{t,q}) }_{\mathrm{scalar}}  
    \end{align*}
    \item Let $C_{p,q} \in \R^{d \times d}$ denote the $(p,q)$-th block of $C$, 
    \begin{align*}
        C_{p,q}= \sum_{i=1}^m \underbrace{ y_i y_i^\top }_{ d \times d \mathrm{~matrix} } \cdot \underbrace{ (x_i^\top u_{t,p} ) }_{ \mathrm{scalar} } \cdot \underbrace{ (x_i^\top u_{*q}) }_{\mathrm{scalar}}
    \end{align*}
    \item Let $D_{p,q} \in \R^{d \times d}$ denote the $(p,q)$-th block of $D$, 
    \begin{align*}
        D_{p,q}= u_{t,p}^\top u_{*q} I 
    \end{align*}
    \item Let $S_{p,q} \in \R^{d \times d}$ denote the $(p,q)$-th block of $S$, 
    \begin{align*}
        S_{p,q}= \begin{cases}
        \sigma_p^* I , & \mathrm{if} ~ p=q; \\
        ~ 0, & \mathrm{if} ~ p \ne q.
        \end{cases}
    \end{align*}
    Here $\sigma_{1}^*, \cdots \sigma_k^*$ are singular values of $W_* \in \R^{d \times d}$.
   \item We define $F \in \R^{d \times k}$ as follows
   \begin{align*}
   \underbrace{ \vect(F) }_{d \times 1} := \underbrace{ B^{-1} }_{d \times d} \underbrace{ (BD-C) }_{d \times d} \underbrace{ S }_{d \times d} \cdot \underbrace{ \vect(V_*) }_{d \times 1}.
   \end{align*}
\end{itemize}
\end{definition}

\subsection{Upper Bound on \texorpdfstring{$\| BD - C \|$}{}}
\label{sec:bdc_upper}
\begin{claim}\label{cla:spectral_BD_minus_C}
Let $B, C$ and $D$ be defined as Definition~\ref{def:B_C_D_S}. Then we have
\begin{align*}
\|BD-C\| \leq \epsilon \cdot {\dist}(U,U_*) \cdot  k
\end{align*}

\end{claim}
\begin{proof}

Let $z_1, \cdots, z_k \in \R^{d}$ denote $k$ vectors. Let $z = \begin{bmatrix} z_1 \\ \vdots \\ z_k \end{bmatrix}$. 

We define $f(z):=z^\top (BD-  C)z$

We define $f(z,p,q) = z_p^\top (BD-C)_{p,q} z_q$.

Then we can rewrite
\begin{align*}
z^\top (BD - C) z
= & ~ \sum_{p=1}^k \sum_{q=1}^k z_p^\top (BD-C)_{p,q} z_q \\
= & ~ \sum_{p=1}^k \sum_{q=1}^k z_p^\top ( B_{p,:} D_{:,q} - C_{p,q} ) z_q  \\
= & ~  \sum_{p=1}^k \sum_{q=1}^k z_p^\top ( \sum_{l=1}^k B_{p,l} D_{l,q} - C_{p,q} ) z_q  \\
\end{align*}
By definition, we know
\begin{align*}
B_{p,l} = & ~ \sum_{i=1}^m y_i y_i^\top (x_i^\top u_{t,p}) \cdot (  u_{t,l}^\top x_i ) \\
D_{l,q} = & ~ (u_{*,q}^\top u_{t,l} ) I_d \\
C_{p,q} = & ~ \sum_{i=1}^m y_i y_i^\top (x_i^\top u_{t,p}) \cdot (  u_{*,q}^\top x_i )
\end{align*}

We can rewrite $C_{p,q}$ as follows
\begin{align}\label{eq:rewrite_C_p_q}
C_{p,q} = \sum_{i=1}^m y_i y_i^\top \cdot (x_i^\top u_{t,p}) \cdot (  u_{*,q}^\top I_d x_i )
\end{align}

Let us compute 
\begin{align*}
B_{p,l} D_{l,q} 
= & ~ \sum_{i=1}^m y_i y_i^\top (x_i^\top u_{t,p}) \cdot ( u_{t,l}^\top x_i  ) \cdot ( u_{*,q}^\top u_{t,l}  )  \\
= & ~ \sum_{i=1}^m y_i y_i^\top (x_i^\top u_{t,p}) \cdot  ( u_{*,q}^\top u_{t,l}  ) \cdot ( u_{t,l}^\top x_i  )  
\end{align*}
where the second step follows from $a \cdot b = b \cdot a$ for any two scalars.

Taking the summation over all $l \in [k]$, we have
\begin{align}\label{eq:sum_l_B_p_l_D_l_q}
\sum_{l=1}^k B_{p,l} D_{l,q} 
= & ~ \sum_{l=1}^k \sum_{i=1}^m y_i y_i^\top (x_i^\top u_{t,p}) \cdot  ( u_{*,q}^\top u_{t,l}  ) \cdot ( u_{t,l}^\top x_i  ) \notag  \\
= & ~ \sum_{i=1}^m y_i y_i^\top (x_i^\top u_{t,p}) \cdot   u_{*,q}^\top  \sum_{l=1}^k (u_{t,l} \cdot u_{t,l}^\top ) x_i   \notag \\
= & ~ \sum_{i=1}^m \underbrace{ y_i y_i^\top }_{ \mathrm{matrix} }  \cdot \underbrace{ (x_i^\top u_{t,p}) }_{\mathrm{scalar}} \cdot \underbrace{  u_{*,q}^\top  U_t U_t^\top x_i }_{\mathrm{scalar} }
\end{align}
where first step follows from definition of $B$ and $D$.

Then, we have
\begin{align*}
\sum_{l=1}^k B_{p,l} D_{l,q} - C_{p,q}
= & ~ (\sum_{i=1}^m \underbrace{ y_i y_i^\top }_{ \mathrm{matrix} }  \cdot \underbrace{ (x_i^\top u_{t,p}) }_{\mathrm{scalar}} \cdot \underbrace{  u_{*,q}^\top  U_t U_t^\top x_i }_{\mathrm{scalar} }) - C_{p,q} \\
= & ~ (\sum_{i=1}^m \underbrace{ y_i y_i^\top }_{ \mathrm{matrix} }  \cdot \underbrace{ (x_i^\top u_{t,p}) }_{\mathrm{scalar}} \cdot \underbrace{  u_{*,q}^\top  U_t U_t^\top x_i }_{\mathrm{scalar} }) - (\sum_{i=1}^m y_i y_i^\top \cdot (x_i^\top u_{t,p}) \cdot (  u_{*,q}^\top I_d x_i )) \\
= & ~ \sum_{i=1}^m \underbrace{ y_i y_i^\top }_{\mathrm{matrix} } \cdot \underbrace{ (x_i^\top u_{t,p}) }_{ \mathrm{scalar} } \cdot \underbrace{  u_{*,q}^\top ( U_t U_t^\top - I_d) x_i }_{ \mathrm{scalar} } 
\end{align*}
where the first step follows from Eq.~\eqref{eq:sum_l_B_p_l_D_l_q}, the second step follows from Eq.~\eqref{eq:rewrite_C_p_q}, the last step follows from merging the terms to obtain $(U_t U_t^\top - I_d)$.

Thus,
\begin{align*}
f(z,p,q)
= & ~z_p^\top ( \sum_{l=1}^k B_{p,l} D_{l,q} - C_{p,q} ) z_q \\
= & ~  \sum_{i=1}^m \underbrace{ ( z_p^\top y_i ) }_{ \mathrm{scalar} } \underbrace{ ( y_i^\top z_q ) }_{\mathrm{scalar}} \cdot \underbrace{ (x_i^\top u_{t,p}) }_{\mathrm{scalar}} \cdot \underbrace{   u_{*,q}^\top  ( U_t U_t^\top - I_d) x_i }_{\mathrm{scalar}}
\end{align*}

For easy of analysis, we define $v_t:= u_{*,q}^\top  ( U_t U_t^\top - I_d)$. This means $v_t$ lies in the complement of span of $U_t$. 

Then 
\begin{align}\label{eq:upper_bound_v_t}
\| v_t \|_2 
= & ~ \| u_{*,q}^\top  ( U_t U_t^\top - I_d) \|_2 \notag \\
= & ~ \| e_q^\top U_*^\top (U_t U_t^\top - I_d) \|  \notag \\
\leq & ~ \| U_*^\top (U_t U_t^\top - I_d) \| \notag \\
= & ~  \dist(U_*,U_t).
\end{align}
where the second step follows from $u_{*,q}^\top = e_q^\top U_*^\top$ ($e_q \in \R^k$ is the vector $q$-th location is $1$ and all other locations are $0$s), 
third step follows from Fact~\ref{fac:norm}.

We want to apply Definition~\ref{def:G_operator}, but the issue is $z_p, z_q$ and $v_t$ are not unit vectors. So normalize them. Let $\ov{z}_p = z_p / \|z_p \|_2$ , $\ov{z}_q = z_q / \|z_q \|_2$ and $\ov{v}_t = v_t/ \| v_t \|_2$.

In order to apply for Definition~\ref{def:G_operator}, we also need $v_t^\top u_{t,p}=0$. 

This is obvious true, since $v_t$ lies in the complement of span of $U_t$ and $u_{t,p}$ in the span of $U_t$. 

We define 
\begin{align*}
G := \sum_{i=1}^m \underbrace{ (x_i^\top u_{t,p}) }_{\mathrm{scalar}} \cdot \underbrace{ (x_i^\top \ov{v}_t) }_{ \mathrm{scalar} }  \cdot \underbrace{ y_i y_i^\top }_{ \mathrm{matrix} }
\end{align*}

By  Definition~\ref{def:G_operator}, we know that 
\begin{align*}
\| G \| \leq \epsilon.
\end{align*}
By definition of spectral norm, we have for any unit vector $\ov{z}_p$ and $\ov{z}_q$, we know that
\begin{align*}
|\ov{z}_p^\top G \ov{z}_q | \leq \| G \| \leq \epsilon.
\end{align*}
where the first step follows from definition of spectral norm (Fact~\ref{fac:norm}), and the last step follows from Definition~\ref{def:G_operator}.

Note that
\begin{align*}
f(p,q,z) = & ~ \sum_{i=1}^m \underbrace{ (x_i^\top u_{t,p}) \cdot (x_i^\top \ov{v}_t) }_{ \mathrm{scalar} } \cdot \underbrace{ (\ov{z}_p^\top y_i) \cdot (y_i^\top \ov{z}_q) }_{ \mathrm{scalar} } \cdot \underbrace{ \| z_p \|_2 \cdot \| z_q \|_2 \cdot \| v_t \|_2 }_{ \mathrm{scalar} } \\
= & ~ \underbrace{ \ov{z}_p^\top }_{ 1 \times d} \cdot \Big( \sum_{i=1}^m \underbrace{ (x_i^\top u_{t,p}) \cdot (x_i^\top \ov{v}_t) }_{ \mathrm{scalar} } \cdot \underbrace{ y_i y_i^\top }_{ d \times d } \Big) \cdot \underbrace{ \ov{z}_q }_{d \times 1} \cdot \underbrace{ \| z_p \|_2 \cdot \| z_q \|_2 \cdot \| v_t \|_2 }_{\mathrm{scalar}} \\
= & ~ \underbrace{ \ov{z}_p^\top }_{1 \times d} \cdot \underbrace{ G }_{d \times d} \cdot \underbrace{ \ov{z}_q }_{d \times 1} \cdot \underbrace{ \| z_p \|_2 \cdot \| z_q \|_2 \cdot \| v_t \|_2 }_{ \mathrm{scalar} }
\end{align*}
where the second step follows from rewrite the second scalar $(\ov{z}_p^\top y_i) (y_i^\top \ov{z}_q) = \ov{z}_p^\top (y_i y_i^\top) \ov{z}_q$, the last step follows from definition of $G$.

Then,
\begin{align*}
|f(z,p,q)|
= & ~ | \sum_{i=1}^m \ov{z}_p^\top G \ov{z}_q | \cdot \| z_p \|_2 \| z_q \|_2 \| v_t \|_2 \\
\leq & ~ \epsilon  \| z_p \|_2 \| z_q \|_2 \cdot \| v_t \|_2 \\
\leq & ~ \epsilon  \| z_p \|_2 \| z_q \|_2 \cdot \dist(U_t,U_*)
\end{align*}
where the last step follows from Eq.~\eqref{eq:upper_bound_v_t}.

Finally, we have
\begin{align}\label{eq:upper_bound_BD_minus_C}
    \|BD-C\|
    = & ~ \max_{z,\|z\|_2=1}|z^\top(BD-C)z| \notag \\
    = & ~ \max_{z,\|z\|_2=1}\big|\sum_{ p \in [ k ],q \in [k] }f(z,p,q)\big| \notag \\
    \leq & ~ \max_{z,\|z\|_2=1}\sum_{ p \in [ k ],q \in [k] }| f(z,p,q)| \notag \\
    \leq & ~ \epsilon \cdot \dist(U_t,U_*) \max_{z,\|z\|_2=1}\sum_{p\in [k] , q \in [k] }  \|z_p\|_2\|z_q\|_2 \notag \\
    \leq & ~ \epsilon \cdot {\dist}(U,U_*) \cdot  k
\end{align}
where the first step follows from Fact~\ref{fac:norm}, the last step step follows from $\sum_{p=1}^k \| z_p \|_2 \leq \sqrt{k} (\sum_{p=1}^k \| z_p \|_2^2)^{1/2} = \sqrt{k}$.

\end{proof}

\subsection{Rewrite \texorpdfstring{$V_{t+1}$}{}}
\label{sec:rewrite_V}

\begin{claim}\label{cla:V_*_bot_dot_V_t+1}
If 
\begin{align*}
 V_{t+1} = (W_*^\top U_t-F)R^{-1}
\end{align*}
then, 
\begin{align*}
(V_{*,\bot} )^\top V_{t+1} = -(V_{*,\bot} )^\top FR^{-1}
\end{align*}
\end{claim}
\begin{proof}

    Multiplying both sides by $V_{*,\bot} \in \R^{d \times (d-k)}$:
    \begin{align*}
        V_{t+1}=&~ (W_*^\top U_t-F)R^{-1}\\
        (V_{*,\bot} )^\top V_{t+1}=&~(V_{*,\bot} )^\top(W_*^\top U_t-F)R^{-1}\\
        (V_{*,\bot} )^\top V_{t+1}=&~(V_{*,\bot} )^\top W_*^\top R^{-1}-(V_{*,\bot} )^\top FR^{-1}
    \end{align*}
    We just need to show $(V_{*,\bot} )^\top W_*^\top R^{-1}=0$.

By definition of $V_{*,\bot}$, we know:
\begin{align*}
    V_{*,\bot}^\top V_*={\bf 0}_{k \times (n-k)}
\end{align*}

Thus, we have:
\begin{align*}
    (V_{*,\bot} )^\top W_*^\top =&~ V_{*,\bot}^\top V_* \Sigma_* U_*^\top \\
    =&~ 0
\end{align*}
 
\end{proof}

\subsection{Upper bound on \texorpdfstring{$\| F \|$}{}}
\label{sec:F_upper}
\begin{lemma}[A variation of Lemma 2 in \cite{zjd15}]\label{lem:upper_bound_norm_F}
Let $\mathcal{A}$ be a rank-one measurement operator where $A_i = x_i u_i^\top$. Let $\kappa$ be defined as Definition~\ref{def:kappa}.

 Then, we have
 \begin{align*}
\| F \| \leq 2 \epsilon k^{1.5} \cdot \sigma_1^* \cdot \dist(U_t,U_*)
 \end{align*}

 Further, if $\epsilon \leq 0.001 / ( k^{1.5} \kappa )$
\begin{align*}
    \| F \| \leq 0.01 \cdot \sigma_k^* \cdot {\dist}(U_t,U_*).
\end{align*}
\end{lemma}
\begin{proof}
Recall that 
\begin{align*}
\vect(F) = B^{-1}(BD-C)S \cdot \vect(V_*).
\end{align*}

Here, we can upper bound $\| F \|$ as follows
\begin{align}\label{eq:upper_bound_norm_F}
    \|F\| \leq & ~ \|F\|_F \notag \\
    = & ~ \| \vect(F) \|_2 \notag \\
    \leq & ~ \|B^{-1}\| \cdot \|BD-C\| \cdot \|S\| \cdot \|\vect(V_*)\|_2 \notag \\
    = & ~ \|B^{-1}\| \cdot \|(BD-C)\| \cdot \| S \| \cdot \sqrt{k} \notag\\
    \leq & ~ \|B^{-1}\| \cdot \|(BD-C)\| \cdot  \sigma_1^* \cdot \sqrt{k}
\end{align}
where the first step follows from $\| \cdot \| \leq \| \cdot \|_F$ (Fact~\ref{fac:norm}), the second step follows vectorization of $F$ is a vector, the third step follows from $\| A x \|_2 \leq \|A \| \cdot \| x \|_2$, the forth step follows from $\| \vect(V_*) \|_2 = \| V_* \|_F \leq \sqrt{k}$  
(Fact~\ref{fac:norm}) and the last step follows from $\| S \| \leq \sigma_1^*$ (see Definition~\ref{def:B_C_D_S}). 

Now, we first bound $\|B^{-1}\|=1/(\sigma_{\min}(B))$. 

Also, let $Z=\begin{bmatrix}z_1 & z_2 & \cdots &  z_k \end{bmatrix}$ and let $z=\vect(Z)$.

Note that $B_{p,q}$ denotes the $(p,q)$-th block of $B$.

We define 
\begin{align*}
{\cal B} := \{ x \in \R^{kd} ~|~ \| x \|_2 = 1 \}.
\end{align*}

Then

\begin{align}\label{eq:variation_min_sigma_B}
    \sigma_{\min}(B)
    = & ~ \min_{z \in {\cal B} }z^\top B z \notag \\
    = & ~ \min_{z \in {\cal B} }\sum_{ p \in [k], q \in [k] }z_p^\top B_{pq}z_q \notag \\
    = & ~ \min_{z \in {\cal B} }\sum_{p=1}^k z_p^\top B_{p,p}z_p+\sum_{p\neq q}z_p^\top B_{p,q}z_q.
\end{align}
where the first step follows from Fact~\ref{fac:norm}, 
the second step follows from simple algebra, the last step follows from  
(Fact~\ref{fac:norm}).

We can lower bound $z_p^\top B_{p,p}z_p$ as follows
\begin{align}\label{eq:lower_bound_sigma_B_pp}
    z_p^\top B_{p,p} z_p
    \geq & ~ \sigma_{\min}(B_{p,p}) \cdot \| z_p \|_2^2 \notag \\
    \geq & ~ (1-\epsilon) \cdot \| z_p \|_2^2
\end{align}
where the first step follows from Fact~\ref{fac:norm}  
, the last step follows from Definition~\ref{def:B_operator} .

We can upper bound $| z^\top B_{p,q} z_q |$ as follows,
\begin{align}\label{eq:upper_bound_norm_B_pq}
    |z_p^\top B_{p,q} z_q|
    \leq & ~ \| z_p \|_2 \cdot \| B_{p,q} \| \cdot \| z_q \|_2 \notag \\
    \leq & ~ \epsilon \cdot \| z_p \|_2 \cdot \| z_q \|_2
\end{align}
where the first step follows from Fact~\ref{fac:norm}, the last step follows from Definition~\ref{def:B_operator} . 

We have
\begin{align}\label{eq:lower_bound_sigma_min_B}
    \sigma_{\min}(B)
    = & ~ \min_{z,\|z\|_2=1}\sum_{p=1}^k z_p^\top B_{p,p}z_p+\sum_{p\neq q}z_p^\top B_{p,q}z_q \notag \\
    \geq & ~ \min_{z,\|z\|_2=1} (1-\epsilon)\sum_{p=1}^k \| z_p \|_2^2 +\sum_{p\neq q}z_p^\top B_{p,q}z_q \notag \\
    \ge & ~ \min_{z,\|z\|_2=1}(1-\epsilon)\sum_{p=1}^k\|z_p\|_2^2-\epsilon\sum_{p \neq q}\|z_p\|_2\|z_q\|_2 \notag \\
    = & ~ \min_{z,\|z\|_2=1} (1-\epsilon) -\epsilon\sum_{p \neq q}\|z_p\|_2\|z_q\|_2  \notag \\
    = & ~ \min_{z,\|z\|_2=1} (1-\epsilon) - k \epsilon \notag \\
    \ge & ~ 1- 2 k\epsilon \notag \\
    \geq & ~ 1/2
\end{align}

where the first step follows from Eq.~\eqref{eq:variation_min_sigma_B},
the second step follows from Eq.~\eqref{eq:lower_bound_sigma_B_pp}, the third step follows from Eq.~\eqref{eq:upper_bound_norm_B_pq}, the forth step follows from $\sum_{p=1}^k \| z_p \|_2^2 = 1$(which derived from the $\|z\|_2=1$ constraint and the definition of $\|z\|_2$), the fifth step follows from $\sum_{p \neq q} \| z_p \|_2 \| z_q \|_2 \leq k$, 
and the last step follows from $\epsilon \leq 0.1/k$. 

 We can show that
 \begin{align}\label{eq:upper_bound_spectral_B_inverse}
\| B^{-1} \| = \sigma_{\min}(B) \leq 2.
 \end{align}
 where the first step follows from Fact~\ref{fac:norm}, the second step follows from Eq.~\eqref{eq:lower_bound_sigma_min_B}.

Now, consider $BD-C$, using Claim~\ref{cla:spectral_BD_minus_C}, we have
\begin{align*}
 \|BD-C\|\leq  k\cdot\epsilon \cdot {\dist}(U_t,U_*)
\end{align*}

Now, we have
\begin{align*}
\| F \| 
\leq & ~ \| B^{-1} \| \cdot \| (BD - C) \| \cdot \sigma_1^* \cdot \sqrt{k} \\
\leq & ~ 2 \cdot \| (BD - C) \| \cdot \sigma_1^* \cdot \sqrt{k} \\
\leq & ~ 2 \cdot k \cdot \epsilon \cdot \dist(U_t,U_*) \cdot \sigma_1^* \cdot \sqrt{k} 
\end{align*} 
where the first step follows from Eq~.\eqref{eq:upper_bound_norm_F}, the second step follows from Eq.~\eqref{eq:upper_bound_spectral_B_inverse},  
and the third step follows from  Eq.~\eqref{eq:upper_bound_BD_minus_C}.
\end{proof}

\subsection{Upper bound on \texorpdfstring{$\| R^{-1} \|$}{}}
\label{sec:R_inv_upper}
\begin{lemma}[A variation of Lemma 3 in \cite{zjd15}]\label{lem:upper_bound_R_inverse_norm}
Let $\mathcal{A}$ be a rank-one measurement operator matrix where $A_i=x_i y_i^\top$. Also, let $\mathcal{A}$ satisfy three properties mentioned in Theorem \ref{thm:main_convergence}.

If the following condition holds
\begin{itemize}
    \item $\dist(U_t, U_*) \leq \frac{1}{4} \leq \epsilon_d = 1/10$ (The condition of Part 1 of Lemma~\ref{lem:main_induction})
\end{itemize}

Then, 
\begin{align*}
    \|R^{-1}\| 
    \leq & ~ 10 /\sigma_{k^*}
\end{align*}
\end{lemma}
\begin{proof}
For simplicity, in the following proof, we use $V$ to denote $V_{t+1}$. We use $U$ to denote $U_t$.

Using Fact~\ref{fac:norm}
\begin{align*}
\| R^{-1} \| = \sigma_{\min}(R)^{-1}
\end{align*}

We can lower bound $\sigma_{\min}(R)$ as follows:
\begin{align}
    \sigma_{\min}(R)
    = & ~ \min_{z,\|z\|_2=1}\|Rz\|_2 \notag \\
    = & ~ \min_{z,\|z\|_2=1}\|VRz\|_2 \notag \\
    = & ~ \min_{z,\|z\|_2=1}\|V_*\Sigma_*U_*^\top Uz-Fz\|_2 \notag \\
    \ge & ~\min_{z,\|z\|_2=1}\|V_*\Sigma_*U_*^\top Uz\|_2-\|Fz\|_2 \notag \\
    \ge & ~\min_{z,\|z\|_2=1}\|V_*\Sigma_*U_*^\top Uz\|_2-\|F\| 
\end{align}
where the first step follows from definition of $\sigma_{\min}$,
the second step follows from Fact~\ref{fac:norm},  
the third step follows from $V = (W_*^\top U-F)R^{-1} =  (V_* \Sigma_* U_*^\top U - F) R^{-1}$ (due to Eq.~\eqref{eq:variation_V} and Definition~\ref{def:W_*})  
, the forth step follows from triangle inequality,  
the fifth step follows from $\| A x \|_2 \leq \| A \|$ for all $\| x \|_2=1$.

Next, we can show that
\begin{align*}
\min_{z,\|z\|_2=1}\|V_*\Sigma_*U_*^\top Uz\|_2
= & ~ \min_{z,\|z\|_2=1}\| \Sigma_*U_*^\top Uz\|_2 \\
\geq & ~ \min_{z,\|z\|_2=1} \sigma_k^* \cdot \| U_*^\top Uz\|_2 \\
= & ~ \sigma_k^* \cdot \sigma_{\min}(U^\top U_*)  
\end{align*} 
where the first step follows from Fact~\ref{fac:norm},  
 the second step follows from Fact~\ref{fac:norm}, the third step follows from definition of $\sigma_{\min}$,  
 
Next, we have
\begin{align*}
\sigma_{\min}(U^\top U_*) 
= & ~ \cos \theta(U_*, U) \\
= & ~ \sqrt{1-\sin^2 \theta(U_*,U)} \\ 
\geq & ~  \sqrt{1- {\dist}(U_*,U)^2}
\end{align*} 
where the first step follows definition of $\cos$, the second step follows from $\sin^2 \theta + \cos^2 \theta  =1$ (Lemma~\ref{lem:sin^2_and_cos^2_is_1}), the third step follows from $\sin \leq \dist$ (see Definition~\ref{def:angle_and_distance}).

Putting it all together, we have
\begin{align*}
\sigma_{\min}(R) \geq & ~ \sigma_k^* \sqrt{1-\dist(U_*,U)^2} - \| F \| \\
\geq & ~ \sigma_k^* \sqrt{1-\dist(U_*,U)^2} - 0.001 \sigma_k^* \dist(U_*,U) \\
= & ~ \sigma_k^* ( \sqrt{1-\dist(U_*,U)^2} - 0.001 \dist(U_*,U)  ) \\
\geq & ~ 0.2 \sigma_k^*
\end{align*}
where the second step follows from Lemma~\ref{lem:upper_bound_norm_F}, the last step follows from $\dist(U_*,U) < 1/10$.

\end{proof}

%% file: reg.tex
\section{Matrix Sensing Regression}
\label{sec:mat_sens_reg}

Our algorithm has $O(\log(1/\epsilon_0))$ iterations, in previous section we have proved why is that number of iterations sufficient. In order to show the final running time, we still need to provide a bound for the time we spend in each iteration. In this section, we prove a bound for cost per iteration. 
In Section~\ref{sec:mat_sens_def_equiv} we provide a basic claim that, our sensing problem is equivalent to some regression problem. In Section~\ref{sec:mat_sens2reg} we show the different running time of the two implementation of each iteration. In Section~\ref{sec:mat_sens_solver} we provide the time analysis for each of the iteration of our solver. In Section~\ref{sec:mat_sens_straigjt_solver} shows the complexity for the straightforward solver. Finally in Section~\ref{sec:mat_sens_cond_no} we show the bound for the condition number. 

\subsection{Definition and Equivalence}
\label{sec:mat_sens_def_equiv}
In matrix sensing, we need to solve the following problem per iteration:

\begin{definition}\label{def:sensing_regression}
Let $A_1,\ldots,A_m \in \R^{d\times d}$, $U\in \R^{d\times k}$ and $b\in \R^m$ be given. The goal is to solve the following minimization problem
\begin{align*}
    \min_{V\in \R^{d\times k}} \sum_{i=1}^m (\tr[A_i^\top U V^\top]-b_i)^2,
\end{align*}
\end{definition}

We define another regression problem 
\begin{definition}\label{def:main_regression}
Let $A_1,\ldots,A_m \in \R^{d\times d}$, $U\in \R^{d\times k}$ and $b\in \R^m$ be given.

We define matrix $M\in \R^{m\times dk}$ as follows
\begin{align*}
    M_{i,*} := & ~ \vect(U^\top A_i), ~~~ \forall i \in [m].
\end{align*}

The goal is to solve the following minimization problem.
\begin{align*}
    \min_{v\in \R^{d k}} \|Mv-b \|_2^2,
\end{align*}
\end{definition}

We can prove the following equivalence result

\begin{lemma}[\cite{z23}]\label{lem:equiv_regression}
Let $A_1,\ldots,A_m \in \R^{d\times d}$, $U\in \R^{d\times k}$ and $b\in \R^m$ be given.

If the following conditions hold
\begin{itemize}
    \item  $
    M_{i,*} :=  \vect(U^\top A_i), ~~~ \forall i \in [m].
$
\item The solution matrix $V \in \R^{d \times k}$ can be reshaped through vector $v \in \R^{dk}$, i.e., $v = \vect(V^\top)$.
\end{itemize}

Then, the problem (defined in Definition~\ref{def:sensing_regression}) is equivalent to problem (defined in Definition~\ref{def:main_regression}) .
 
\end{lemma}

\begin{proof}
Let $X, Y\in \R^{d\times d}$, we want to show that 
\begin{align}
\label{eq:trace_vector}
    \tr[X^\top Y] = & ~ \vect(X)^\top \vect(Y).
\end{align}
Note that the RHS is essentially
$\sum_{i \in [d]}\sum_{j \in [d]} X_{i,j}Y_{i,j}$, for the LHS, note that 
\begin{align*}
    (X^\top Y)_{j,j} = & ~ \sum_{i\in [d]} X_{i,j} Y_{i,j}, 
\end{align*}
the trace is then sum over $j$. 

Thus, we have Eq.~\eqref{eq:trace_vector}. This means that for each $i\in [d]$, 
\begin{align*}
    \tr[A_i^\top UV^\top]=\vect(U^\top A_i)^\top \vect(V^\top).
\end{align*}

Set $M\in \R^{m \times dk}$ be the matrix where each row is $\vect(U^\top A_i)$, we see Definition~\ref{def:sensing_regression} is equivalent to solve the regression problem as in the statement. This completes the proof.
\end{proof}

\subsection{From Sensing Matrix to Regression Matrix}
\label{sec:mat_sens2reg}

\begin{definition}\label{def:M}
Let $A_1,\ldots,A_m \in \R^{d\times d}$, $U\in \R^{d\times k}$ . 
 We define matrix $M\in \R^{m\times dk}$ as follows
\begin{align*}
    M_{i,*} := & ~ \vect(U^\top A_i), ~~~ \forall i \in [m].
\end{align*}
\end{definition}
\begin{claim}\label{cla:time_naive_M}
The naive implementation of computing $M \in \R^{m \times dk}$
takes $m \cdot \Tmat(k,d,d)$ time.
Without using fast matrix multiplication, it is $O(md^2k)$ time.
\end{claim}
\begin{proof}
For each $i \in [m]$, computing matrix $U^\top \in \R^{k \times d}$ times $A_i \in \R^{d \times d}$ takes $\Tmat(k,d,d)$ time. Thus, we complete the proof.
\end{proof}

\begin{claim}\label{cla:time_batch_M}
The batch implementation takes $\Tmat(k,dm,d)$ time. 
Without using fast matrix multiplication, it takes $O(md^2 k)$ time.
\end{claim}
\begin{proof}
We can stack all the $A_i$ together, then we matrix multiplication. For example, we construct matrix $A \in \R^{d \times dm}$. Then computing $U^\top A$ takes $\Tmat(k,d,dm)$ time.
\end{proof}
The above two approach only has difference when we use fast matrix multiplication.

\subsection{Our Fast Regression Solver}
In this section, we provide the results of our fast regression solver. Our approach is basically as in \cite{gsyz23}. For detailed analysis, we refer the readers to the Section~5 in \cite{gsyz23}. 
\label{sec:mat_sens_solver}
\begin{lemma}[Main Cost Per Iteration]\label{lem:fast_solver}
Assume $m = \Omega(dk)$.  
There is an algorithm that runs in time
\begin{align*}
\wt{O}( m d^2 k + d^3 k^3 )
\end{align*}
and outputs a $v'$ such that 
\begin{align*}
\| M v' - b \|_2 \leq (1+\epsilon) \min_{v \in \R^{dk}} \| M v - b\|_2 
\end{align*}
\end{lemma}

\begin{proof}
From Claim~\ref{cla:time_batch_M}, writing down $M \in \R^{m \times dk}$ takes $O(md^2 k)$ time.

Using Fast regression resolver as \cite{gsyz23}, the fast regression solver takes
 
\begin{align*}
O( ( m\cdot dk  + (dk)^3 ) \cdot \log(\kappa(M)/\epsilon) \cdot \log^2(n/\delta) )
\end{align*}
\end{proof}

\begin{lemma}[Formal version of Theorem~\ref{thm:main_informal}]\label{thm:main_cost}
In each iteration, our requires takes $\wt{O}( m d^2 k)$ time.
\end{lemma}
\begin{proof}
Finally, in order to run Lemma~\ref{lem:fast_solver}, we need to argue that $\kappa(M) \leq \poly(k,d,\kappa(W_*))$.

 This is true because $\kappa(U) \leq O(\kappa(W_*))$ and condition number of random Gaussian matrices is bounded by $\poly(k,d)$.

Then applying Lemma~\ref{lem:kappa_M}, we can bound $\kappa(M)$ in each iteration.

Eventually, we just run standard error analysis in \cite{gsyz23}. Thus, we should get the desired speedup.

The reason we can drop the $(dk)^3$ is $m \geq dk^2$.
\end{proof}

\subsection{Straightforward Solver}
\label{sec:mat_sens_straigjt_solver}
Note that from sample complexity analysis, we know that $m = \Omega(dk)$.
\begin{lemma}
Assume $m = \Omega(dk)$. 
The straightforward implementation of the regression problem (Defintion~\ref{def:main_regression}) takes 
\begin{align*} 
O(md^2 k^2)
\end{align*}
time.
\end{lemma}
\begin{proof}
The algorithm has two steps. From Claim~\ref{cla:time_batch_M}, writing down $M \in \R^{m \times dk}$ takes $O(md^2 k)$ time.

The first step is writing down the matrix $M \in \R^{m \times dk}$.

The second step is solving regression, it needs to compute $M^{\dagger} b$ (where $M^{\dagger} \in \R^{d k \times m}$ )
\begin{align*}
 M^\dagger b = ( M^\top M )^{-1} M b
\end{align*}

this will take time
\begin{align*}
\Tmat(dk,m,dk) + \Tmat(dk,dk,dk) 
= & ~ m d^2k^2 + (dk)^3 \\
= & ~ m d^2 k^2
\end{align*}
the second step follows from $m =\Omega(dk)$
.

Thus, the total time is
\begin{align*}
m d^2 k + md^2 k^2 = O(m d^2k^2)
\end{align*}
\end{proof}

\subsection{Condition Number}
\label{sec:mat_sens_cond_no}

\begin{lemma}\label{lem:kappa_M}
We define $B \in \R^{m \times k}$ as follows $B := X U$ and $X \in \R^{m \times d}$ and $U \in \R^{d \times k}$.

Then, we can rewrite $M \in \R^{m \times dk}$
 
\begin{align*}
 \underbrace{M}_{m \times dk} = \underbrace{B}_{m \times k} \otimes \underbrace{Y}_{m \times d}
\end{align*}

Then, we know that $\kappa(M) = \kappa(B) \cdot \kappa(Y) \leq \kappa(U) \kappa(X) \kappa(Y)$.
 
\end{lemma}

\begin{proof}

Recall $U \in \R^{d \times k}$. Then we define $b_i = U^\top x_i$ for each $i \in [m]$.

Then we have
\begin{align*}
M_{i,*} = \vect( U^\top x_i y_i^\top ) = \vect(b_i y_i^\top ).
\end{align*}

Thus, it implies 
\begin{align*}
 M = B \otimes Y
\end{align*}

\end{proof}